\documentclass[11pt]{article}
\usepackage{graphicx}
\usepackage{epstopdf}
\usepackage{amsmath}
\usepackage{amssymb}
\usepackage{verbatim}
\usepackage{algorithmic}
\usepackage{subfig}
\begin{document}
\title{Monte Carlo simulation of the effects of higher order anisotropy on the spin reorientation transition in the two dimensional Heisenberg model with long range interactions }
\author{R. L.~Stamps$^{1,2}$ and M. C.~Ambrose $^{1}$}
\maketitle
\noindent $^{1}$ School of Physics, The University of Western Australia, 35 Stirling Hwy, Crawley 6009, Australia
\\ $^{2}$ SUPA School of Physics and Astronomy, University of Glasgow, Glasgow G12 8QQ, United Kingdom
\date{\today}
\section*{Abstract}
pacs: 75.30.Kz,64.70.dj,07.05.Tp,75.30.Cr
\\

The strength of perpendicular anisotropy is known to drive the spin reorientation in thin magnetic films. Here we consider the effect different order anisotropies have on two phase transitions; the spin reorientation transition and the orientational order transition. We find that the relative magnitude of different order anisotropies can significantly enhance or suppress the degree to which the system reorients. Specifically Monte Carlo simulations reveal significant changes in the cone angle and planar magnetization. In order to facilitate rapid computation we have developed a stream processing technique, suitable for use on GPU systems, for computing the transition probabilities in two dimensional systems with dipole interactions. 
\section{Introduction}
In two dimensions the effect of thermal fluctuations is enhanced. The number of possible symmetries is lower than in three dimensional systems and this reduced symmetry means that there are fewer degrees of freedom to absorb energy \cite{LandauOrigin,PeierlsOrigin,mermin1966absence}. The discovery of a divergence in the susceptibility in the XY model \cite{PhysRevLett.17.913}, caused by topological excitations \cite{0022-3719-6-7-010,ISI:A1971I949100031}, has meant that the existence and stability of spontaneous ordered states in two dimensions has been a rich and often contentious  \cite{PhysRev.127.359, wegner1967spin,PhysRevLett.17.913,PhysRev.158.383,PhysRev.176.250,mermin1966absence,%
PhysRevB.50.3052,springerlink:10.1007/BF02894837,Polyakov197579,%
springerlink:10.1023/A:1013726826390,1751-8121-40-14-001} area of theoretical interest. In particular the dimensionality of the field\cite{1751-8121-40-14-001,PhysRev.65.117}, finite size effects \cite{PhysRevB.60.11761} and anisotropies \cite{0953-8984-20-1-015208} can effect the phase diagram.\\
Two dimensional systems can be realized experimentally as thin (typically $<$ 15 atomic layer) magnetic films. 
By varying the composition and thickness of thin films a large variety of magnetic properties have been obtained\cite{Jensen2006129}. It is possible to create films which strongly favor either in-plane or perpendicular orientation of the magnetization \cite{PhysRevLett.56.2728}. The functional dependence of energy on perpendicular magnetization varies and can be altered using ion beam irradiation\cite{PhysRevB.62.5794,PhysRevB.85.054427}. In the presence of strong uni-axial anisotropy favoring perpendicular alignment, the competition between the entropy favored in-plane magnetization and the energetically favored out of plane state can lead to a temperature driven spin reorientation transition \cite{Politi1995647,PhysRevLett.69.3385, PhysRevLett.64.3179}. The nature of this transition is known to be dependent on the relative strengths of different orders of anisotropy \cite{PhysRevB.55.3708,PhysRevB.51.12563,0295-5075-39-5-557}. In these systems the ratio of long range dipole coupling and short range exchange coupling can lead to the formation of striped domains of alternating spin direction; either as the ground state \cite{rossol:5263, PhysRevLett.84.2247} or as a mechanism of spin reversal \cite{PhysRevB.68.212404}. Films have been produced in with stripes observed running parallel with a common orientation\cite{PhysRevB.68.212404, PhysRevLett.84.2247,PhysRevLett.93.117205}, as zig-zags between regular defects \cite{konings:054306} or forming complex patterns with no orientational order \cite{PhysRevB.55.2752, PhysRevLett.84.2247}. The width and mobility of stripes depends on temperature \cite{PhysRevB.83.172406}. In particular these can systems display strong thermal memory, in which the domain configuration depend on the rate of heating or cooling \cite{PhysRevB.80.184412,PhysRevB.84.094428}.
Both the reorientation and stripe melting transitions have been studied analytically  \cite{PhysRevB.42.849,PhysRevLett.65.2599, PhysRevB.48.10335,PhysRevB.51.1023} and using Monte Carlo Simulation \cite{PhysRevB.77.134417, PhysRevB.77.174415}.\\
Here we propose a technique in which energy differences arising from long range dipole coupling are approximated. This approximation allows for computation to be parallelized significantly; reducing the computational time required. Having first examined the extent to which this approximation influences the results of simulation, the method is applied to a two dimensional Heisenberg model where the effects of higher order anisotropy are examined. 
\section{Theory}
\subsection{Dipole Coupling}
\label{DPC}
Consider a thin ferromagnet, modeled as a two dimensional square lattice of Heisenberg spins ($\vec{s} \in \mathcal{S}^2$). The spins experience a long range dipole interaction 
\begin{equation}
H_{D} = \frac{C_D}{2} \sum_{n,m} \frac{1}{r_{nm}^3} (\vec{s}_n \cdot \vec{s}_m -3 \vec{s}_n \cdot \hat{r}_{nm}\vec{s}_m \cdot \hat{r}_{nm} ).
\end{equation}
 Where $n$ and  $m$ represent vertices of the two dimensional lattice, $\vec{r}_{nm}$ is the vector from $n$ to $m$ and $C_D$ is a constant, $C_D= (M^2 \mu_0)/(4 \pi )$.\\
 In simulations periodic boundary conditions are used to approximate an infinite system. The total system consists of a tiling of replicas. For simulation size $L \times L$, spins separated by vector $\vec{G}= (a L, b L) $ with $a, b \in \mathcal{Z}$ are identical. To compute the infinite sum introduced by periodic boundary conditions a new set of coordinates is introduced; $\vec{r}_{nm} = \vec{G}+\vec{\rho}_{nm}$. Here $\vec{\rho}$ restricted to $\vec{\rho} = (\rho_x,\rho_y)$, with $\rho_x,\rho_y \in [0,L]$. The dipole energy at a site $n$ can then be written (taking the square lattice to be in the $x-y$ plane), 
\begin{equation}
\begin{split}
H_n =& \frac{1}{2} C_D \sum_{m \neq n} \sum_{\vec{G}}  s_n^\alpha s_m^\beta \lim_{r \rightarrow 0}\partial _\alpha \partial _\beta \frac{1}{\left | \vec{\rho}_{nm}+\vec{G}-\vec{r} \right|}
\\
& + C_D   (s_n^\alpha)^2 \lim_{r \rightarrow 0}\partial^2 _\alpha \sum_{\vec{G} \neq 0} \frac{1}{\left | \vec{G}-\vec{r} \right|} ,
\end{split}
\end{equation}
where $s_n^\alpha$ represents $\vec{s}_n.\hat{\alpha}$ and expressions are summed over repeated Greek indexes. In order to achieve efficient computation $ \sum_{\vec{G}} \lim_{r \rightarrow 0}\partial _\alpha \partial _\beta \frac{1}{\left | \vec{\rho}_{nm}+\vec{G}-\vec{r} \right|}$ can be calculated in advance for all choices of $n$, $m$, $\alpha$ and $\beta$. Since the sum is slow to converge it can be split into a short range real space term and a long range Fourier space term according to the technique described by Harris \cite{Harris} based on the analogous three dimensional case developed by Ewald \cite{Ewald}. Letting $f(\vec{r})= \sum_{\vec{G}} \frac{1}{\left | -\vec{\rho}_{nm}+\vec{G}+\vec{r} \right|}$ one has $f= f_L + f_S$ \footnote{While our results take place in the $x-y$ plane the general results given below contain $z_{nm}$ and are suitable for three dimensional systems that are infinite in two directions.}
\begin{equation}
f_S(\vec{r}) =\sum_{\vec{G}}\frac{1}{|\vec{r}-\vec{\rho}_{nm}+\vec{G}|} \text{erfc}\left(\frac{|\vec{r}-\vec{\rho}_{nm}+\vec{G}|}{2 \eta}\right)
\end{equation}
and
\begin{equation}
f_L(\vec{r}) =1/L^2 \sum_{\vec{k}}  \tilde{h}^L(\vec{k},z) \exp(i 2 \pi \vec{k} \cdot \vec{r})
\end{equation}
where 
\begin{equation}
\begin{split}
\tilde{h}^L(\vec{k},z) = \frac{\pi}{k} e^{-i \vec{k} \cdot \vec{\rho}_{nm}} [
& e^{k|z-z_{nm}|} \text{erfc}\left( \frac{|z-z_{nm}|}{2 \eta} +k \eta  \right) 
\\
 +&e^{-k|z-z_{nm}|} \text{erfc}\left( \frac{-|z-z_{nm}|}{2 \eta} +k \eta  \right) ]
\end{split}
\end{equation} 
Despite the efficiency gained by pre-calculating the interactions and using this rapid summation technique, the calculation of the dipole interaction is still computationally intensive. In order to calculate the energy of a single spin in the system one must calculate $N=L^2$ interactions, when calculating the energy of a state of the system $C^{N}_{2}$ interactions are required. For a moderate system size $L=64$ this equates to $4096$ interactions for a single spin and over $8 \times 10^6$ for a single state.
\subsection{Monte Carlo Simulation}
When considering a system at finite temperature observable quantities $O$ are calculated as expectation values of the Boltzmann distribution: 
\begin{equation}
\label{expect}
\langle O \rangle = \frac{\sum_{i} \hat{O}[\phi_i] \exp \left(\frac{-H_i}{k_B T}\right) }{Z}
\end{equation}
In order to approximate the properties of this distribution a subset of possible states is selected using a Markov chain Monte Carlo with transition function
\begin{equation}
\label{Metro}
P_T(\phi_j \rightarrow \phi_k ) = \begin{cases}
\exp \left( \frac{-(H_k-H_j)}{k_B T} \right) & \text{for $(H_k-H_j) > 0 $} \\
  1 & \text{otherwise}
\end{cases}
\end{equation}
Known as the metropolis algorithm  \cite{OrigionalMC3}, Eqn. (\ref{Metro}) does not define a method for selecting the prospective new state. There are numerous methods for constructing new states and the decision is based largely on the system being analyzed. In the case of magnetic systems, the simplest and most common choice is single spin flips.
For a system of size $N$ one Monte Carlo step (MC step) requires $N$ spin flips. Herein lies the computational difficulty, in order to complete one MC step in a two dimensional system one must calculate the energy of a single spin $N=L^2$ times. If there is dipole coupling present each energy calculation requires $L^2$ interactions to be computed. In order to compute MC steps more efficiently we present a stream processing algorithm in section \ref{GPU} that  reduces this computational load.
\section{GPU Parallel Programming and Simultaneous flipping}
\label{GPU}
In order to perform Monte Carlo simulations at an acceptable speed we make use of a graphics processing unit (GPU). Unlike most modern computing systems that implement a Harvard execution model, GPU computing employs a stream processing model. In stream processing a single function (kernel) is executed simultaneously on a large number of different inputs (the stream). Importantly the execution of each input (thread) is independent and there is no communication between threads \cite{Nickolls:2008:SPP:1365490.1365500}. For spin lattice models where interactions are limited to nearest or next nearest neighbor the problem of implementing a parallel GPU algorithm has been examined previously and several algorithms exist to distribute computation over single \cite{Weigel201192,Weigel20111833,Preis20094468} or multiple GPUs \cite{Block20101549}. In these cases multiple single spin flips can be performed simultaneously (provided potential update sites are not nearest neighbors). In recent work, Campos et al. \cite{campos} have approached the problem of long range coupling by parallelizing the long range sum in three dimensions.  Here we go beyond the work of Campos et al. and focus on performing multiple simultaneous MC steps. 
\subsection{Algorithm}
\begin{figure}[!htb]
  \centering
  \includegraphics[width=9cm]{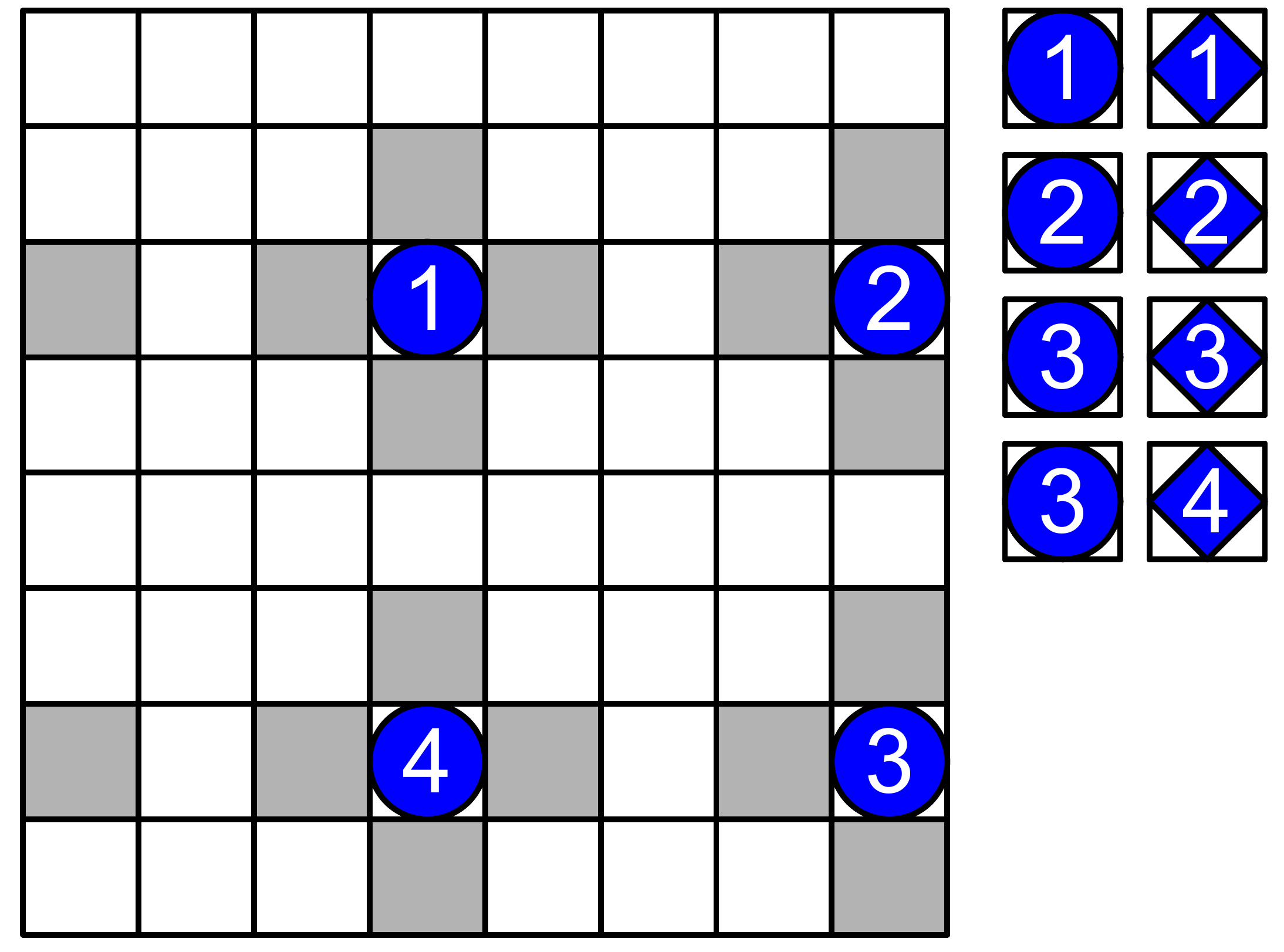}
  \caption[Simultaneous update sites]%
  {An $8 \times 8$ sample where in each square represents a spin site. Site 1 is chosen at random. Update sites 2,3 and 4 are selected at fixed distances from 1. In addition 4 alternative spin values are also generated (indicated here as diamonds) giving a total of $(2L^2)/l^2 = 8$ energies that need to be calculated. Gray squares indicate nearest neighbor sites used in calculation of short range interactions.}
  \label{algo}
\end{figure}

\begin{figure}[!htb]
  \centering
  \includegraphics[width=9cm]{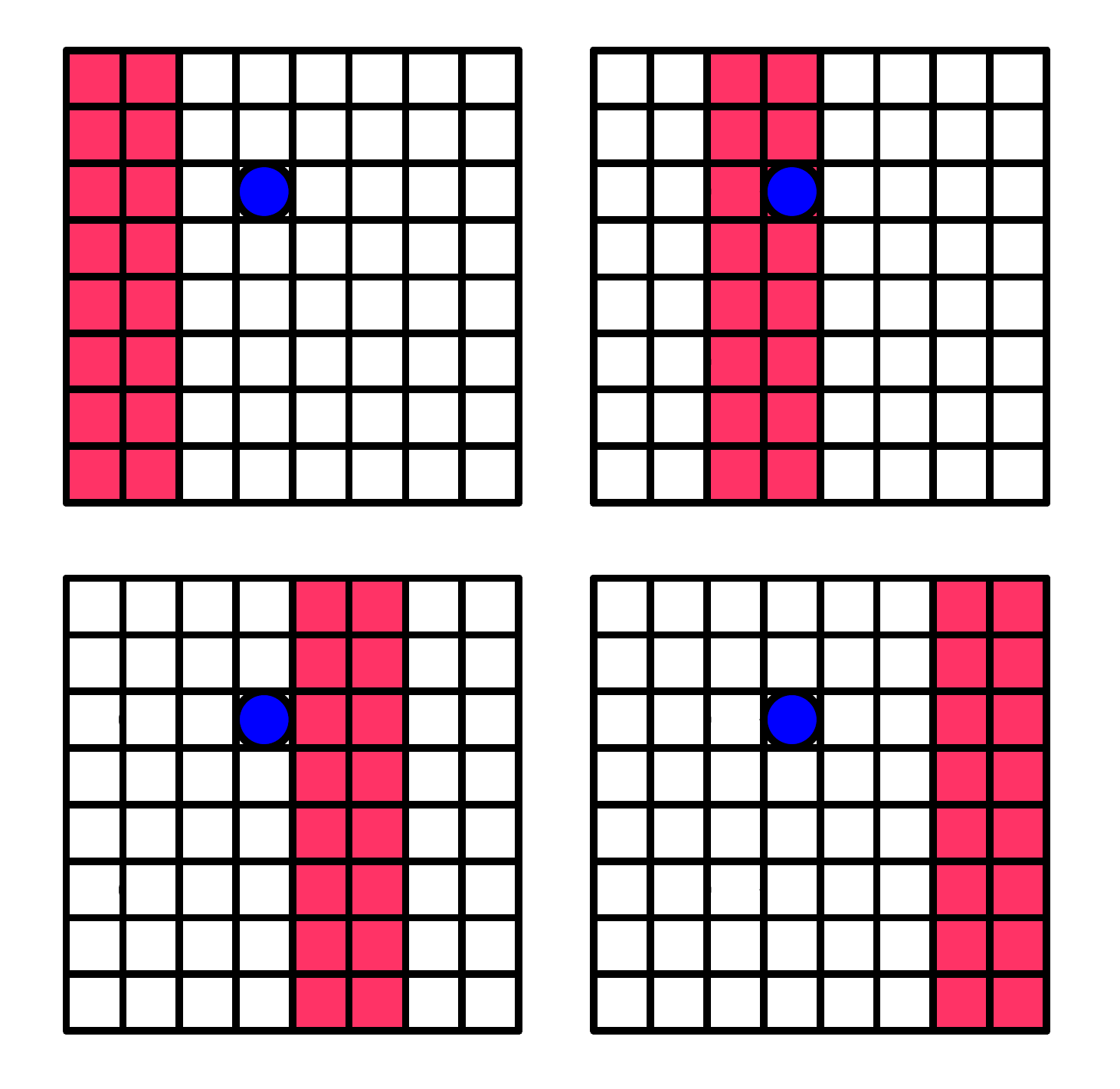}
  \caption[Subsystems of th dipole interaction]%
  {The dipole interactions between update site 1 (blue circle) and the rest of the system are parallelized into four threads which compute a sub set of the possible interactions shown here as pink shaded squares. Simultaneously threads will be executing the calculation of dipole coupling for the other seven spin values.
 }
  \label{algo2}
\end{figure}

\begin{figure}[!htb]
  \centering
  \includegraphics[width=9cm]{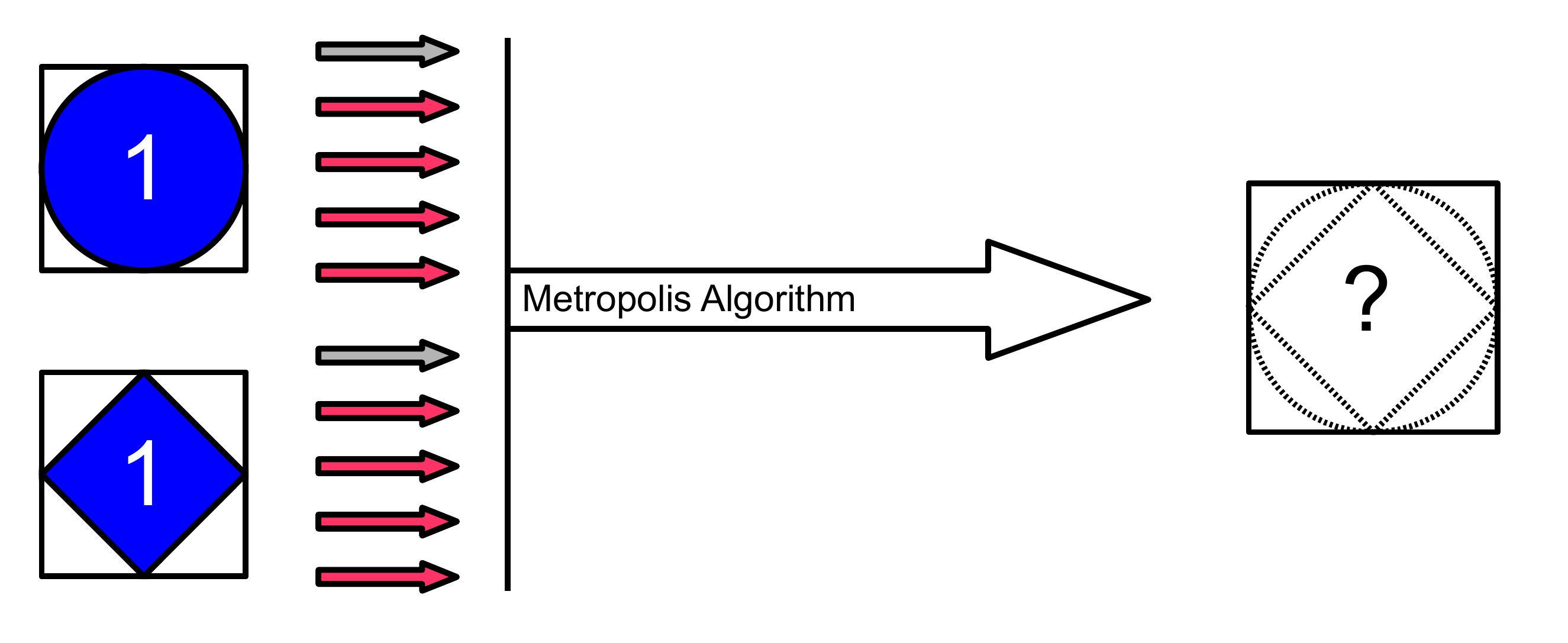}
  \caption[GPU metropolis algorithm]%
  {One of the $n$ simultaneous spin flipping threads. The original spin value and the alternative spin value for site 1 each require five threads to compute all interactions. One thread to calculate the short range interactions with gray squares in Fig. \ref{algo} (represented by a gray arrow) and $P=4$ threads to calculate the dipole interactions with the subsystems shown as pink squares in Fig. \ref{algo2} (represented by pink arrows). The results of the ten threads are then fed into a single thread that calculates the flipping probability for update site 1.
 }
  \label{algo3}
\end{figure}

\begin{figure}[!htb]
  \centering
  \includegraphics[width=9cm]{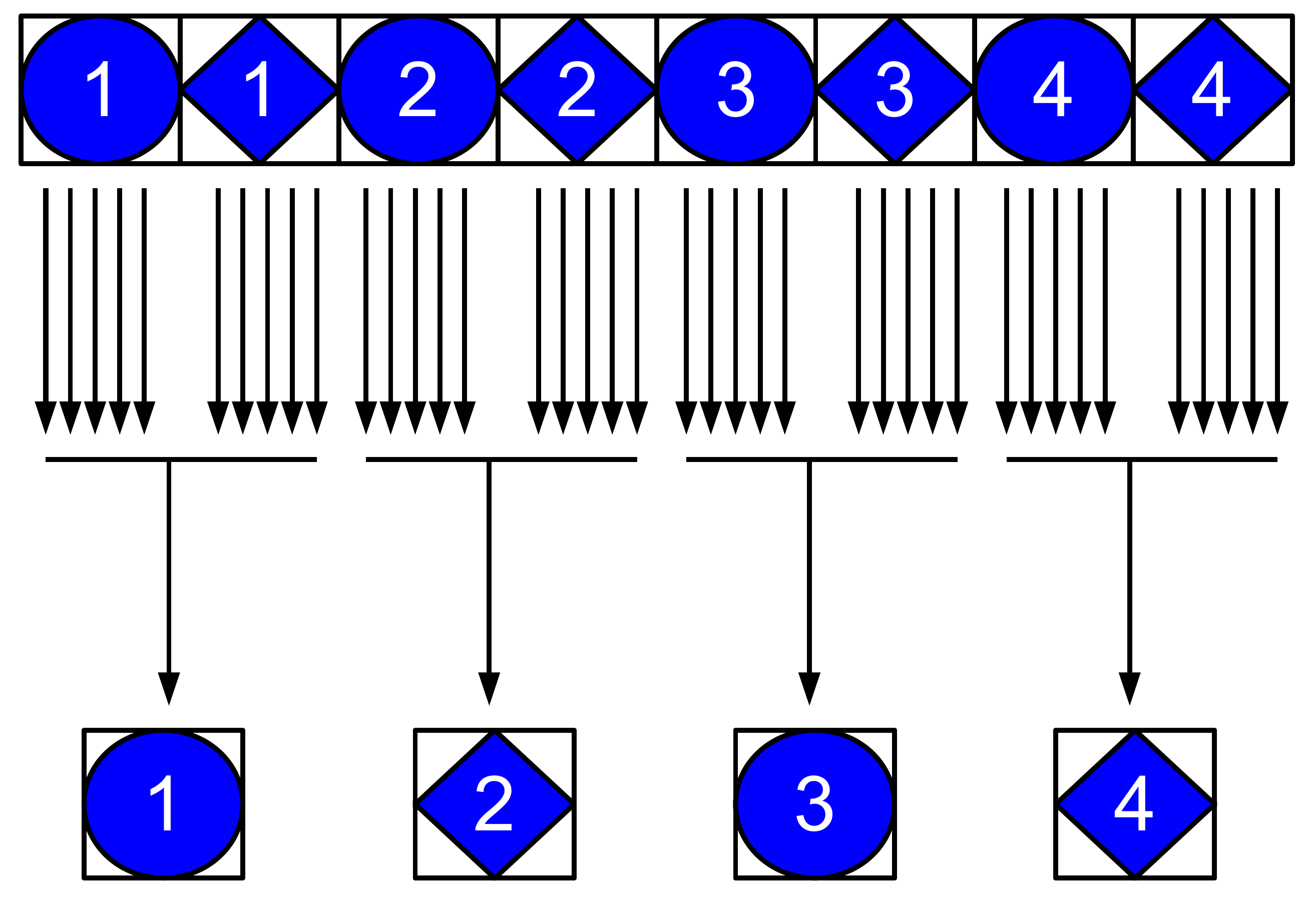}
  \caption[All threads]%
  {For each of the four current spins and each potential new spin the five threads shown in Fig. \ref{algo3} are computed the results are fed into $n=4$ threads that calculate the flipping probability. One of the possible $2^n$ possible results spin updates is shown.
 }
  \label{algo4}
\end{figure}

Here we describe a method for parallelizing MC simulations in the presence of dipole coupling. A pseudo code implementation of the algorithm is given in appendix \ref{Code}. The algorithm depends on the size of the system $L$ and two parameters that will be defined below: $l$ and $P$. For clarity of exposition, figures in this section will use a fixed small system size $L=8$ and the parameters $l = 4$ and $P=4$ ($l$ and $P$ need not be equal in general). When describing the algorithm we take the `host' to indicate any computation not performed on the GPU. Calculations run on host are implemented in the normal serial fashion. We will refer to the GPU card as the `device' and calculations run on device are implemented as parallel operations and have access to device memory (VRAM).\\
Initially the current state of the system is held in the device memory, either from initialization or from the previous iteration of the algorithm. On the host, a site is selected at random, in Fig. \ref{algo} this site is denoted by a blue circle numbered $1$. Additional sites are then selected at fixed multiples of $l$ according to  $i = (al ,bl)+ i_1$ for $a,b \in [0,L/l]$. These $n=(L/l)^2$ values of $i$ are the update sites and are represented in Fig. \ref{algo} as circles numbered 2, 3 and 4. Next for each site a new spin value is selected at random as the potential new spin values. In Figs. \ref{algo}, \ref{algo3} and \ref{algo4} these potential new spin values are represented as diamonds. The location of the selected sites and the potential new spin values are copied to the device memory. 
\\
 The device then launches $2n$ threads to calculate nearest neighbor exchange coupling\footnote{Two threads per update site, to calculate the current energy and the energy of the potential new spins}. The nearest neighbor spins accessed by these threads are indicated as gray boxes in Fig.  \ref{algo}.
Simultaneously the system launches $2Pn$ threads to calculate the dipole interactions. Each thread calculates the interaction between a spin at an update site and spins in a vertical sub-section of the total system with width $L/P$. Fig. \ref{algo2} presents a subset of these threads that calculate the interactions associated with a single update site.
When the above threads are complete all interactions have been computed. For each of the $n$ potential update sites the results of both the current spin and the potential new spin are passed into a single thread. Each of the $n$ new threads calculates any single site energies (anisotropies and applied fields), then applies the metropolis algorithm and updates the state accordingly( see Fig \ref{algo3}).
In Fig. \ref{algo4} all the threads executed in one iteration of the algorithm for the hypothetical small system are displayed.
\subsection{Approximation}
\label{SectA}
The algorithm presented here reduces the computation required by simultaneously executing $P \times (L/l)$ partial sums of size $L/P$. However in doing so an approximation has been made.
\begin{figure}[!htb]
  \centering
  \includegraphics[width=9cm]{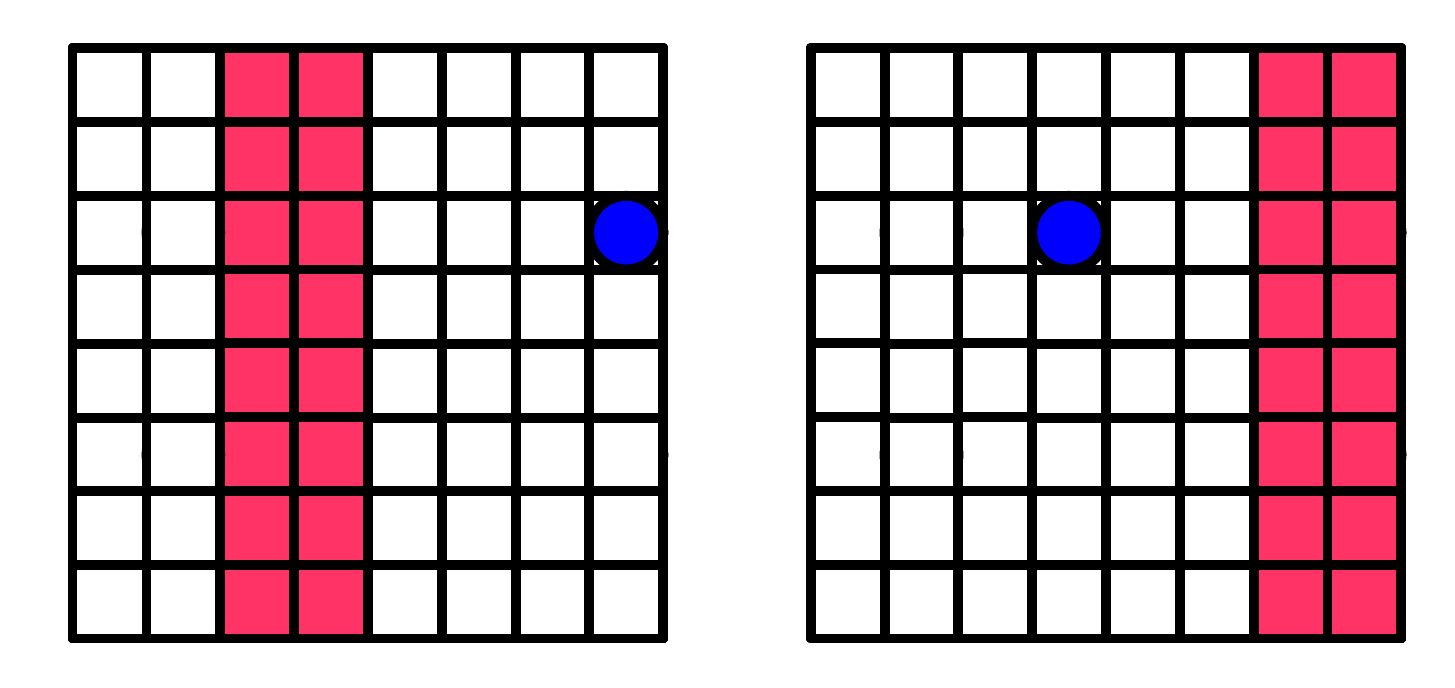}
  \caption[Simultaneous Flips]%
  {The interactions calculated by two of simultaneously executed threads. On the left site 2 interacts with a section of the system that contains site 1. On the right site 1 interacts with a section of the system that contains site 2.}
  \label{algo5}
\end{figure}
In Fig. \ref{algo5}, a single thread will compute the dipole interaction between a potential update site (blue circles in the figure) and a subset of the sample of width $L/P$ (pink). Some subsets will contain another potential update site (the interactions of which are being processed simultaneously another thread).
If two spin updates were computed without parallelization, there are four possibilities for the new state $\phi_i$ given in Fig. \ref{algo6}: neither spin is flipped ($\phi_0$), the first spin flips ($\phi_1$), the second spin flips  ($\phi_2$) or both spins flip  ($\phi_3$). Denote the probability that the first spin flips as $P_1$, the probability that the second spin flips given the first spin has flipped as $P_{12}$ and that the second flips given the first spin is not flipped as $P_{\bar{1}2}$ . Then the probability that system finishes in state $\phi_0$, is  $\mathcal{P}_0 = \neg P_1 \neg P_{\bar{1}2}$, where we have used $\neg P_1$ to denote the probability that $P_1$ does not occur ($\neg P_1 = (1-P_1)$). Similarly $\mathcal{P}_1 = P_1 \neg P_{12}$,  $\mathcal{P}_2 = \neg P_1  P_{\bar{1}2}$ and $\mathcal{P}_3 = \neg P_1  P_{12}$. If the two spins don't interact then $P_{12}=P_{\bar{1}2}$ and depends only on the energies $H_0$ and  $H_2$. In this case simultaneously updating sites gives the same statistics as the conventional sequential flipping. If the spins do interact, as is the case with long range coupling, then the above algorithm makes the approximation $P_{12}= P_T(\phi_1 \rightarrow \phi_3 )  \approx P_T(\phi_0 \rightarrow \phi_2 )$ or $H_3-H_1\approx H_2-H_0$.
\begin{figure}[!htb]
  \centering
  \includegraphics[width=6cm]{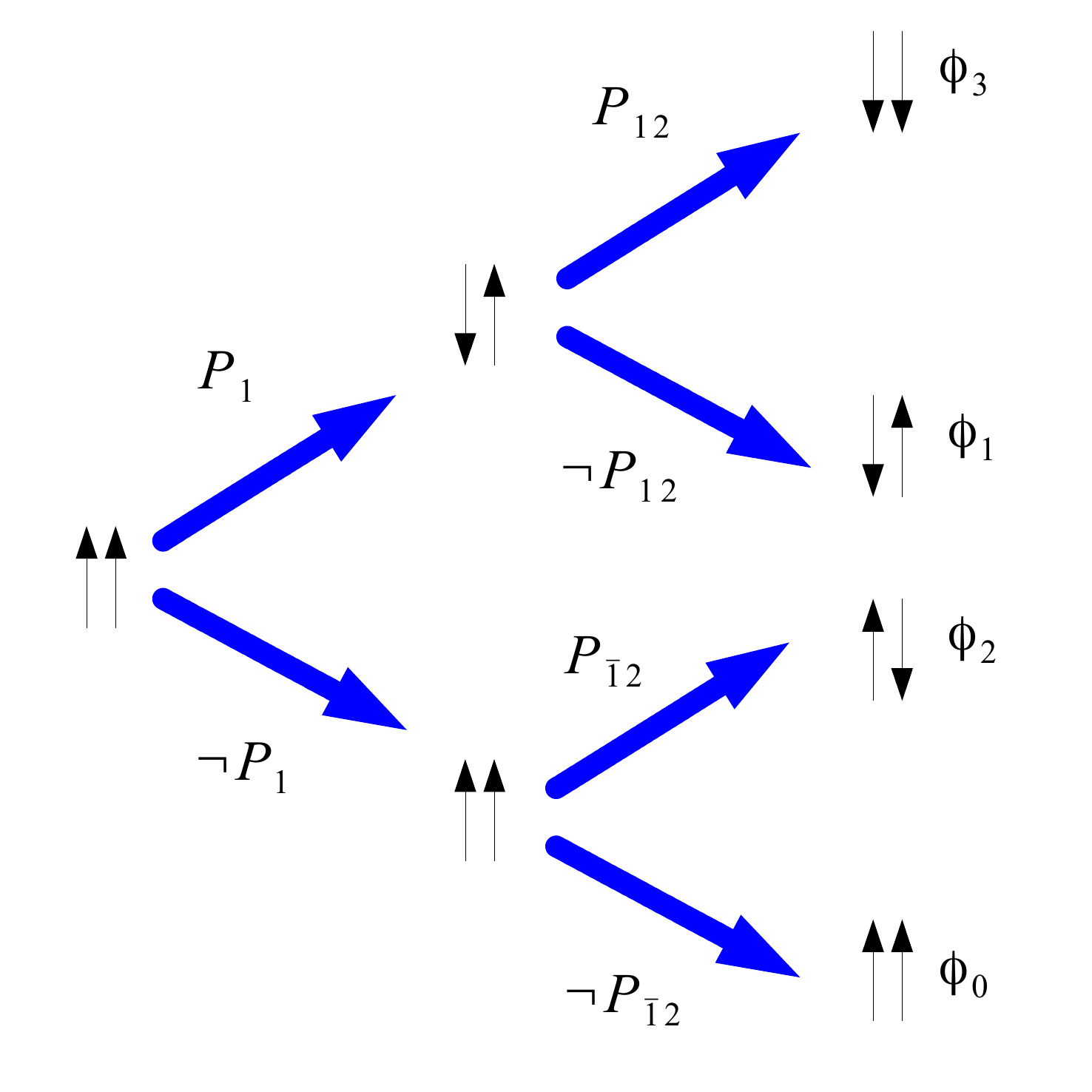}
  \caption[Potential Update sites]%
  {Possible out comes of two potential spin flips and the associated probabilities.}
  \label{algo6}
\end{figure}
\\ In order to approximate the size of this error first denote original spin values as $\vec{s}_1$ and $\vec{s}_2$ and the potential new values as $\vec{S}_1$ and $\vec{S}_2$. Then $\phi_0 = (\vec{s}_1,\vec{s}_2)$, $\phi_1 = (\vec{S}_1,\vec{s}_2)$ , $\phi_2 = (\vec{s}_1,\vec{S}_2)$ and $\phi_3 = (\vec{S}_1,\vec{S}_2)$. Assuming that spins are not nearest neighbors then the error $\epsilon_{12}= |H_3-H_1-H_2+H_0|$ depends only on the dipole energy. The error is 
\begin{equation}
\begin{split}
\epsilon_{12} =&\frac{C_D}{r_{12}^3} |\delta_1\cdot \delta_2 -3  (\delta_1 \cdot \hat{r}_{12})( \delta_2 \cdot \hat{r}_{12}) | \\
=& \frac{C_D}{r_{12}^3} \mathcal{D}.
\end{split}
\end{equation}
where  $\delta_1=\vec{s}_1-\vec{S}_1$, $\delta_2=\vec{s}_2-\vec{S}_2$ and  $\mathcal{D} = |\delta_1 \cdot \delta_2 -3 (\delta_1 \cdot \hat{r}_{12})( \delta_2 \cdot \hat{r}_{12}) | $. We wish to find the maximal values of $\mathcal{D}$. There are two cases in which local maximums occur. The first case, which we shall refer to as the perpendicular case, is $\delta_1= \pm \delta_2 = \pm (0,0,2) $ corresponding to $\mathcal{D} = 2$. The second case, which we shall refer to as the planar case, is $\delta_1= \pm \delta_2 = \pm 2 \hat{r}_{12} $ corresponding to $\mathcal{D} = 4$. \\
Here we have calculated the maximum error introduced by flipping two spins simultaneously. However, due to the periodic boundaries conditions described in section \ref {DPC} flipping a spin also flips its image in each replica that makes up the infinite system. For the perpendicular case each replica introduces an additional error $ \epsilon_{R}=(2C_D)/(|\vec{G}+\vec{r}_{12}|^3) $. For the in-plane case the maximum $\mathcal{D}=4$ requires $\delta_1$ and $\delta_2$ to be parallel to the vector connecting the updated spins, $\vec{r}_{12}+\vec{G}$, which will not be possible for all $\vec{G}$. To estimate the error for the in-plane case let 
\begin{equation}
\frac{\vec{r}_{12}+\vec{G}}{|\vec{r}_{12}+\vec{G}|} =(\cos(\gamma), \sin(\gamma),0)
\end{equation}
 then fixing the $\delta_1 \parallel \delta_i$ gives $\mathcal{D} = 1-3\cos(\gamma)^2$. Since simultaneously flipped spins and their replicas will exist in all possible directions we take the root mean squared average with respect to $\gamma$ giving $\mathcal{D} \approx 2$. Based on this argument we estimate the maximum error in $H_3-H_1$ as
\begin{equation}
\epsilon_{\text{Total}} = \sum_{i=-n}^{n} \sum_{j=-n}^{n} \sum_G  \frac{2 C_D}{\left((il+G^x)^2 + (jl+G^y)^2 \right)^{\frac{3}{2}}}
\end{equation}
The flipping probability will now be $\exp \left( \frac{-(H_k-H_j)}{k_B T}  \pm \frac{\epsilon_{\text{Total}}}{k_B T}\right)$. At high $T$ the Boltzmann probability is more uniform and the error is reduced. For $L=64$, $l=32$ and $C_D^{-1} k_B T = 0.1$ the error in the flipping probability is approximately $0.14\%$. This error places an upper bound on the error for a single spin flip, however, it does not preclude the possibility of large errors accumulated over the many millions of spin flips that will be performed in simulation. In Appendix \ref{DIM} we present the results of simulating the two dimensional Ising model with strong dipole interactions. We find that, when compared with conventional techniques, the algorithm produces a maximum error of around $8\%$ on the obtained critical values.   

\section{Fourth order Anisotropies in the Heisenberg Model}\label{Res}
In two dimensional magnetic systems the phase diagram depends on two ratios. The first is the ratio of perpendicular anisotropy to dipole coupling. In two dimensions the dipole energy favors in-plane ordering of spins at $T=0$ \cite{PhysRevB.77.134417,PhysRevB.77.174415}, however a sufficiently strong perpendicular anisotropy can create a ground state with perpendicular spins. At finite temperature the free energy $F=E-TS$ is minimized and the higher entropy in-plane state can become favored and the spin reorientation transition occurs. The other ratio determining the possible phases of the system is exchange to dipole coupling. For sufficiently strong dipole coupling the system forms stripes in the ground state, with stronger dipole coupling favoring thinner stripes. The total energy for such a system is given by 
\begin{equation}
\begin{split}
 H= &\frac{J}{2}\sum_{\langle i,j \rangle} \vec{s}_i \cdot \vec{s}_j +\sum_i H_{A\,i} \\
 +&\frac{C_D}{2} \sum_{i,j} \frac{1}{r_{ij}^3} (\vec{s}_i \cdot \vec{s}_j -3 \vec{s}_i\cdot\hat{r}_{ij}\vec{s}_j \cdot \hat{r}_{ij} )
\end{split}
\end{equation}
Where $H_{A\,i}$ is the single site magnetic anisotropy. In the absence of stripes the spin reorientation transition is known to depend on the higher order anisotropy terms \cite{Politi1995647,PhysRevLett.69.3385, PhysRevLett.64.3179}. We are interested on the effect of varying the ratio of second to fourth order anisotropy while keeping the anisotropic energy difference between the in-plane and out of plane spins constant. The anisotropy is defined as
\begin{equation}
H_{A\,i} = K( (1-a) (\vec{s}_i \cdot \hat{z})^2 + a (\vec{s}_i \cdot \hat{z})^4)
\end{equation}
with $K<0$ and $a>-1$. Here $K$ represents the strength of the anisotropy and $a$ determines the ratio of fourth order to second order anisotropies. Theoretically the dependence of spin orientation on higher order anisotropy has been studied as a function of thickness \cite{PhysRevB.49.15665} and temperature \cite{PhysRevB.54.4137, PhysRevB.62.14259}. In these cases reorientation of spins is modeled as a competition between competing anisotropy terms which depend on temperature. In our simulations anisotropy is considered constant.\\
Previously the case of $a=0$ and varying $K$ has been examined by Whitehead et al. \cite{PhysRevB.77.174415}. The broad thermal phase evolution, ordered stripes at low temperature followed by in-plane magnetization followed by the paramagnetic transition, is reasonably well understood \cite{PhysRevB.51.1023, PhysRevB.77.174415}. The behavior near to the SRT is not well understood, experiments performed on Pt/Co(0.5 nm)/Pt films by Bergeard et al. \cite{bergeard} have indicated long time scale dynamics consisting of a regions fluctuating between perpendicular stripes and in plane magnetic order.   The authors note that near to the SRT, quadratic coupling alone is not sufficient to account for this mixed behavior. Using AC susceptibility studies of striped phases in Fe/Ni films by Abu-Libdeh et al. \cite{PhysRevB.80.184412,PhysRevB.84.094428,PhysRevB.81.195416}, have also indicated the presence of long time scale dynamics. By varying the order of anisotropy we wish to investigate the nature of the phase transition between the striped and in plane phases. 
\\
Here increasing the value of $a$ suppresses states with canting. We consider three choices of parameter $a$; $a = -1$ corresponding to a system that favors canting (F), $a=2$ corresponding to a state with suppressed canting (S) and $a=0$ corresponding to an intermediate propensity for canting (I).
\\
We understand this as follows. Consider the restoring force due to anisotropy experienced by a spin slightly canted away from perpendicular alignment. For the intermediate case the restoring force is given by the derivative of energy with respect to zenith angle $-\partial_{\theta_i }H_{A_i} |_{\theta_i = 0} = -2 K$ (with the same results for $\theta_i = \pi$), so for a small amount of canting away from the perpendicular alignment  the change in energy is $\Delta H_{A_i}(\theta_i) =2 K \theta_i$.  For the case of canting suppression we have $-\partial_{\theta_i }H_{A_i} |_{\theta_i = 0} = 0$ and so there is no force experienced for small spin canting. For the case of strong suppression $-\partial_{\theta_i }H_{A_i} |_{\theta_i = 0} = -6 K$ and the restoring force is three times stronger than the equivalent quadratic anisotropy.
\subsection{Results}
\begin{figure}[htb]
  \centering
  \includegraphics[width=8cm]{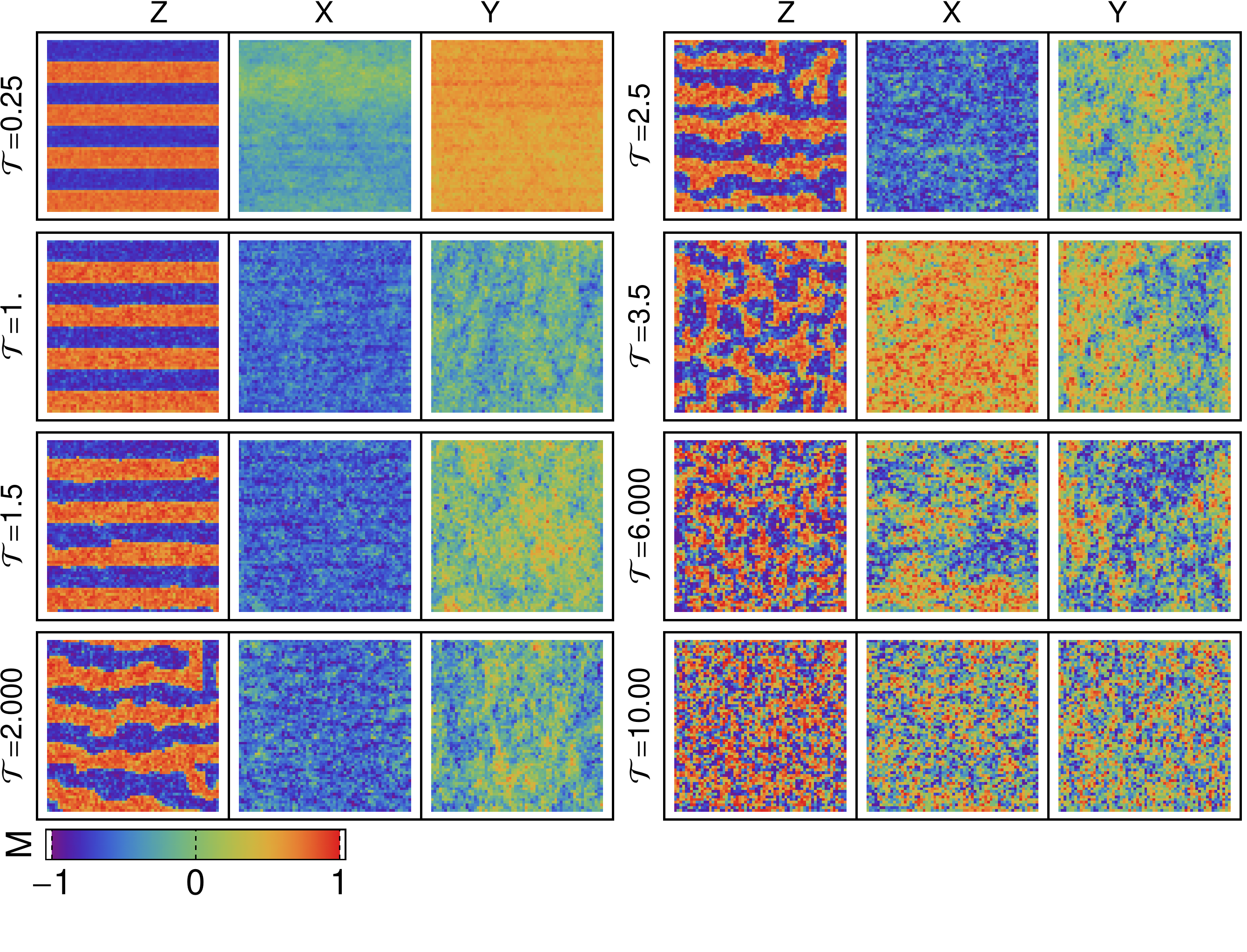}
  \caption[Example of $a=-1$ states]%
  {Example spin configurations for canting favored system($a=-1$). Each state is represented by three images, from left to right, showing the $z$,$x$ and $y$ components of spins respectively. Left column from top $\mathcal{T} = 0.25, 1., 1.5, 2.0$. Right column from top $\mathcal{T}= 2.5, 3.5, 6, 10$.}
   \label{HAM1}
\end{figure}

\begin{figure}[htb]
  \centering
  \includegraphics[width=8cm]{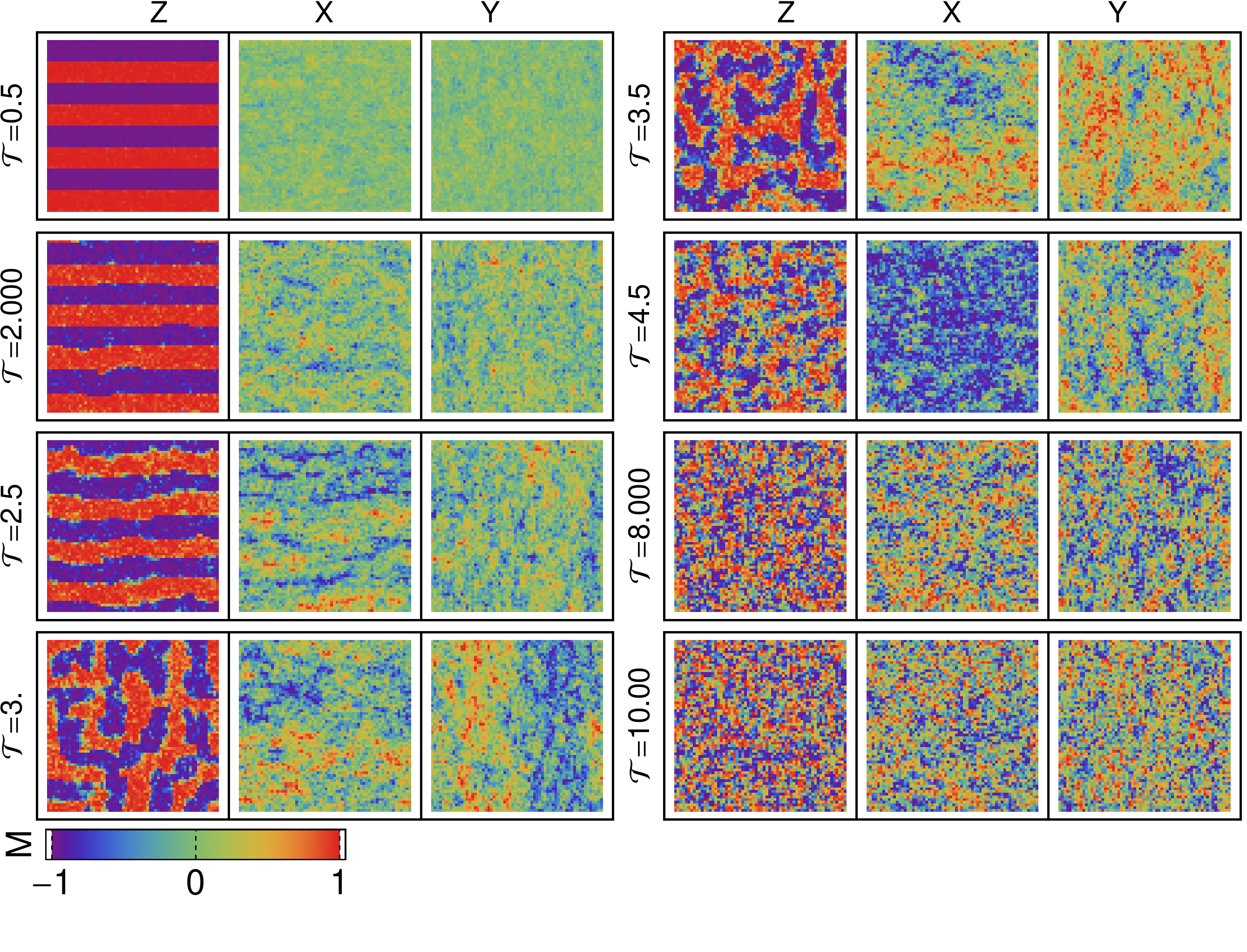}
  \caption[Example of $a=0$ states]
  {Example spin configurations for the intermediate case ($a=0$). Left column from top $\mathcal{T} = 0.5, 2, 2.5, 3$. Right column from top $\mathcal{T}=  3.5, 4.5, 8, 10$.}
  \label{NoHA}
\end{figure}

\begin{figure}[htb]
  \centering
  \includegraphics[width=8cm]{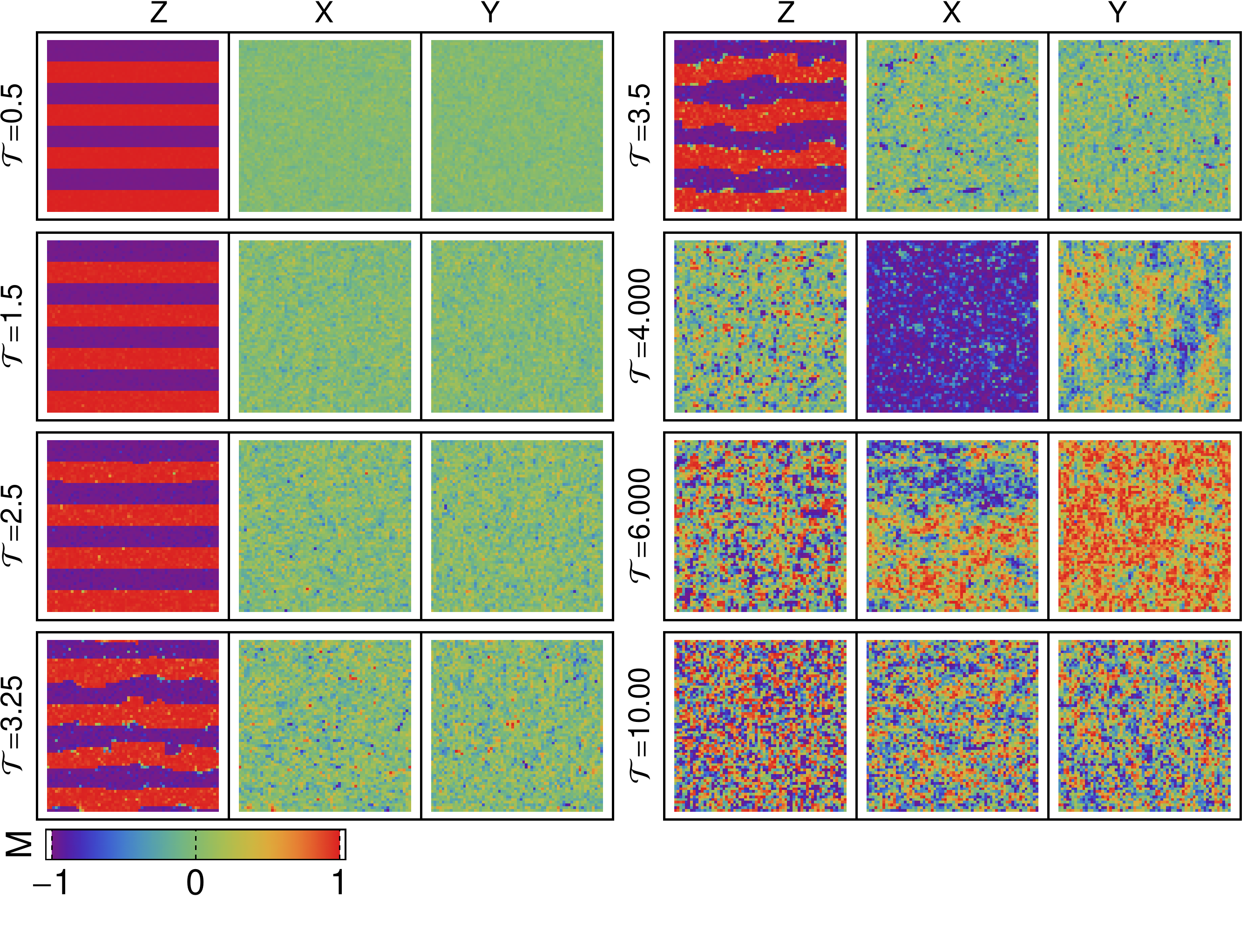}
  \caption[Example of $a=2$ states]%
  {Example spin configurations for the canting suppressed case ($a=2$). Left column from top $\mathcal{T} = 0.5, 1.5, 2.5, 3.25$. Right column from top $\mathcal{T}= 3.5, 4, 6, 10$.}
  \label{HA2}
\end{figure}

\begin{figure}[htb]
  \centering
  \includegraphics[width=6cm]{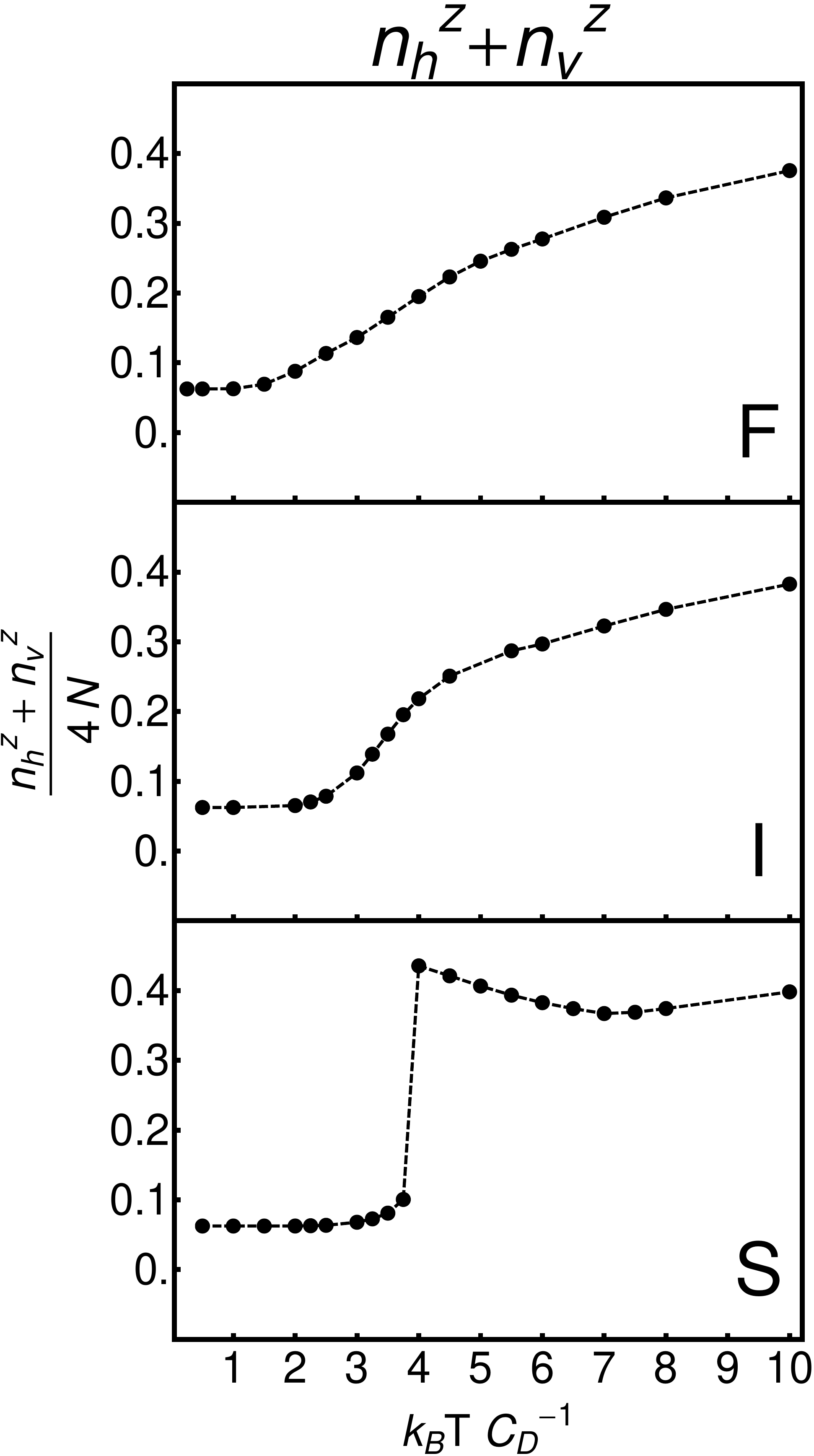}
  \caption[Wall Proportion]%
  {Total wall length for the canting favored $a=-1$ (F), intermediate $a=0$ (I) and canting suppressed $a=2$ (S) cases.}
  \label{WT}
\end{figure}

Normalizing against the strength of the dipole coupling $C_D$ to give dimensionless parameters we define $\mathcal{T}=(k_B T)/C_D$, $\mathcal{J}=J/C_D$ and $\mathcal{K}=K/C_D$. The exchange coupling was fixed at $\mathcal{J}=8.9$ giving a ground state with stripe width $w=8$. The anisotropy strength was fixed at $\mathcal{K}=15$.\\
The system is initialized in a perpendicular striped state and then an ensemble is generated using the above parallel Monte Carlo algorithm with parameters  $L=64$, $P=64$ and $l=32$. While the energy of system converges rapidly (typically taking several thousand MC steps), the morphological properties of the system can take longer to emerge. Slow relaxation of stripe patterns has been observed in studies of Ising systems \cite{Bromley200314}. We find that after an equilibrium time of $10^5$ Monte Carlo steps the order parameters we measure no longer have a time dependence. Once at equilibrium an additional $5 \times 10^4$ steps are simulated with the state recorded every $50$ steps to form an ensemble. This ensures that the correlation between the same spin in subsequent states of the ensemble is limited. Waiting $50$ steps corresponds to an average value of the time correlation $ \langle \vec{s}_i(t=t_o) \cdot \vec{s}_i(t=t_o+50) \rangle \approx 0.5$ at $\mathcal{T} =4$. The ensemble size $n=1000$ ensures that the standard error in the thermal averages of order parameter $O$ 
\begin{equation}
\text{SE}_{\langle O \rangle} =\left (\frac{ \langle O^2 \rangle - \langle O \rangle^2 }{n} \right)^{1/2}
\end{equation}
remains smaller than the errors due to the algorithm discussed in section \ref{GPU}.
Example states for each value of $a$ are given in figures \ref{HAM1}, \ref{NoHA} and \ref{HA2}.
\\
For the case of favored spin canting, $a=-1$, (Fig. \ref{HAM1}) the system has stable stripes in the ground state, but the perpendicular magnetization is not saturated. The in-plane components display long range ordering. As temperature is increased the stripes display roughening, and then bridging leading to the eventual loss of orientational order. As temperature is further increased the in-plane order breaks into domains, before the system enters the high temperature paramagnetic phase.
\\
For the intermediate case, $a=0$ (Fig. \ref{NoHA}) we observe perpendicular stripes at low $\mathcal{T}$. The boundaries of these stripes undergo roughening and eventual bridging, analogous to the canting favored case, before orientational order is destroyed with increasing temperature. Unlike the canting favored case we note the presence of increased canting at the domain boundaries. Above this temperature we observe a mixed phase in which perpendicular domains are interspersed with regions of in-plane magnetic order. As temperature is further increased these domains become increasingly granular until system reaches the paramagnetic limit.
\\
When canting is suppressed, $a=2$, (Fig. \ref{HA2}) the system forms perpendicular stripes in the ground state. As the temperature is increased the walls undergo roughening but not the bridging and gradual loss of orientational order displayed in the $a=0$ and $a=-1$ simulations. Instead the system undergoes a sudden transition into a state with only small regions of perpendicular alignment remaining and strong in-plane order. As temperature is increased we observe an increasing number perpendicular regions as the in-plane order breaks into domains. At high temperature in-plane order is destroyed as the system becomes paramagnetic. 
\\
In order to locate domain walls we define horizontal and vertical order parameters based on those described by Whitehead et al. \cite{PhysRevB.77.174415, PhysRevLett.75.950}
\begin{equation}
\label{HVO}
\begin{split}
n_v^z &=\sum_i 1-\text{sgn}(s_i^z s_{i+\hat{y}} ^z) \\
n_h^z &=\sum_i 1-\text{sgn}(s_i^z s_{i+\hat{x}} ^z)
\end{split}
\end{equation}
with analogous definitions for the $x$ and $y$ components of the spins. At low temperatures when the systems are dominated by perpendicular alignment the density of perpendicular domain boundaries is given by $(n_v^z+n_h^z)/(4N)$. In Fig \ref{WT} we see that for the intermediate and canting favored states the wall density increases gradually with temperature. This represents roughening and then the bridging before perpendicular order is lost and the wall density tends to the $T \rightarrow \infty$ value of $1/2$. In contrast the canting suppressed state undergoes far less roughening and the wall density remains small until the system undergoes a sudden change into the high $T$ state.
\subsubsection{Orientational Order Transition}
\begin{figure}[!htb]
  \centering
  \includegraphics[width=6cm]{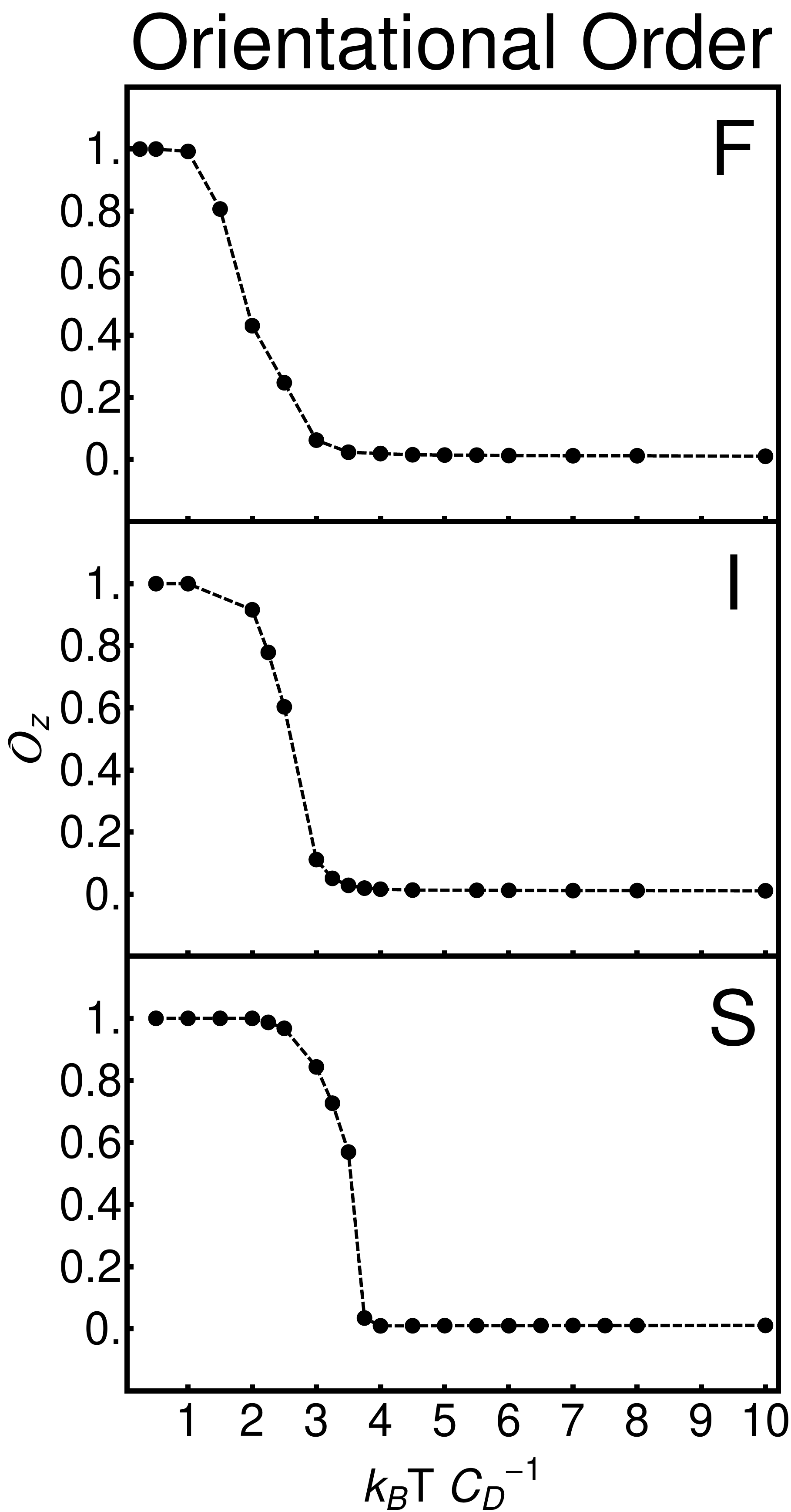}
  \caption[orientational order paramter]%
  {The orientational order for favored canting $a=-1$ (F), intermediate canting $a=0$ (I) and suppressed canting $a=2$ (S).}
  \label{OT}
\end{figure}

The orientational order is given by considering the ratio of horizontal and vertical domain boundaries. 
\begin{equation}
\mathcal{O}_\alpha = \left \langle \frac{|n_h^\alpha-n_v^\alpha |}{n_h^\alpha+n_v^\alpha} \right \rangle
\end{equation}
for $\alpha \in \{x,y,z\}$. In Eq. \ref{HVO} we have defined separate order parameters for each component rather than defining  combined parameters $n_v =\sum_i 1-\vec{s}_i \vec{s}_{i+\hat{y}}$ and  $n_h =\sum_i 1-\vec{s}_i \vec{s}_{i+\hat{x}}$. This definition is robust against canting, meaning reduction in $\mathcal{O}$ due to changing stripe morphology (such as coarsening or bending) can be distinguished from changes in the cone angle. Consider the low $\mathcal{T}$ states shown in Fig. \ref{HAM1} and \ref{HA2}, both states have stripes with no coarsening or bending and the definition given in Eq. \ref{HVO} assigns the same order parameters to both states. In Fig. \ref{OT} this $\mathcal{O}_z$ is plotted for the three values of $a$. $\mathcal{O}_x$ and $\mathcal{O}_y$ were also calculated and were zero in all cases.
 For higher $a$ the stripes are stabilized at higher $\mathcal{T}$. For perpendicular stripes, high $a$ values suppress small fluctuations at the stripe boundaries preventing the roughening and eventual bridging that leads to a loss of orientational order.
\subsubsection{Spin Reorientation Transition}
When the spins are not entirely perpendicular to the plane it is possible for the system to acquire in-plane ferromagnetic order. If we define $M_x = 1/N \sum_i s_i^x$ and $M_y = 1/N \sum_i s_i^y$ then the parallel magnetization is given by
\begin{equation}
M_{\parallel} = \langle(M_x^2+M_y^2)^{\frac{1}{2}} \rangle .
\end{equation}
the appearance of non-zero ferromagnetic order is not inherently indicative of global spin reorientation. Canted spins states and finite thickness domain walls can account for significant magnetic order \cite{PhysRevB.77.174415,PhysRevB.38.9145}. To measure the degree to which the spins reorient as a function of temperature we introduce the cone angle $\eta$
\begin{equation}
\eta =\frac{1}{N} \sum_i\sqrt{ (2/\pi)^{2} \langle(\theta_i-\pi/2)^2 \rangle}
\end{equation}
where $\theta_i$ is the zenith angle of spin $\vec{s}_i$, $\theta_i = \text{arccos}(s_i^z)$ and we have normalized $\eta$ so that $0 \leq \eta  \leq 1$.
\begin{figure}[!htb]
  \centering
  \includegraphics[width=6cm]{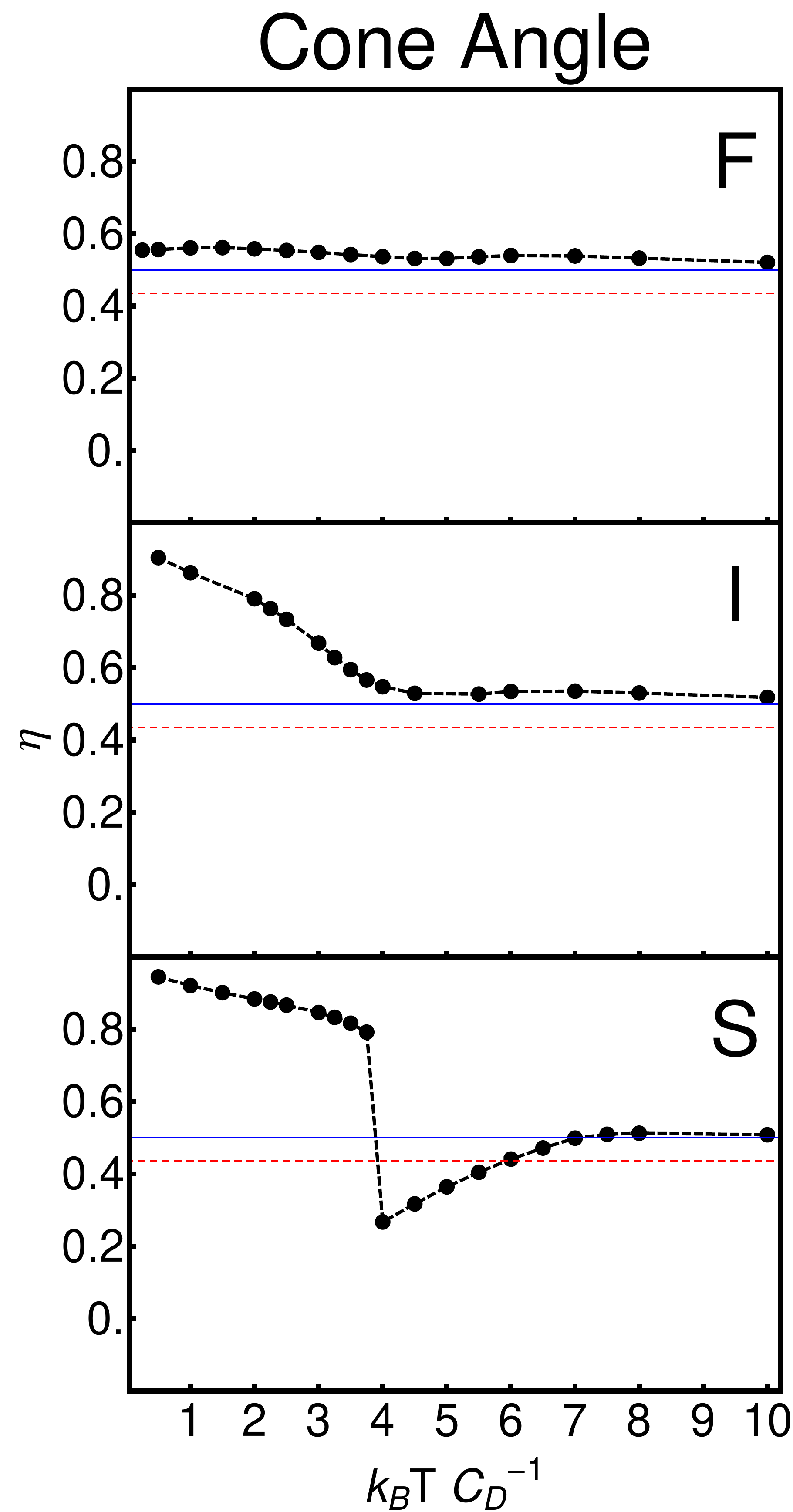}
  \caption[$\eta$ paramter]%
  {$\eta$ as a function of $\mathcal{T}$ for favored canting $a=-1$ (F), intermediate canting $a=0$ (I) and suppressed canting $a=2$ (S). $\eta=1$ indicates all spins point perpendicular to the plane ($\vec{s}_i.\hat{z} =\pm 1$), $\eta=0$ indicates all spins lie in-plane  ($\vec{s}_i.\hat{z} =0$). The red dashed line represents the $T \rightarrow \infty$ value. The solid blue line represents ($\vec{s}_i.\hat{z} = 1 / \sqrt{2}$).}
  \label{CA}
\end{figure}
In Fig. \ref{CA} $\eta$ is shown as a function of $\mathcal{T}$, along with the $T \rightarrow \infty$ value $\eta = \sqrt{ (2/\pi)^{2} (1/4(\pi^2-8))}$. Despite the energy difference between parallel and perpendicular alignments remaining constant, the behavior varies dramatically with $a$. \\
For the canting favored case, $a=-1$, we see that by reducing the energy cost of canting spins, the sample doesn't experience a spin reorientation transition, instead it remains canted for all temperatures. In Fig. \ref{MAG} it is seen that this canting allows the sample to have non zero magnetic order at $\mathcal{T}=0$. The appearance of maximum magnetic ordering at $\mathcal{T}=0$ is characteristic of weaker values of $\mathcal{K} < 13$ and $a=0$ \cite{PhysRevB.77.174415}. However in these cases, the low $\mathcal{T}$ parallel magnetization does not occur simultaneously with orientational order. As $\mathcal{T}$ is increased the degree of canting remains practically unchanged with the orientational order decreasing over the range of $1 < \mathcal{T}<2.5$.
\begin{figure}[!htb]
  \centering
  \includegraphics[width=6cm]{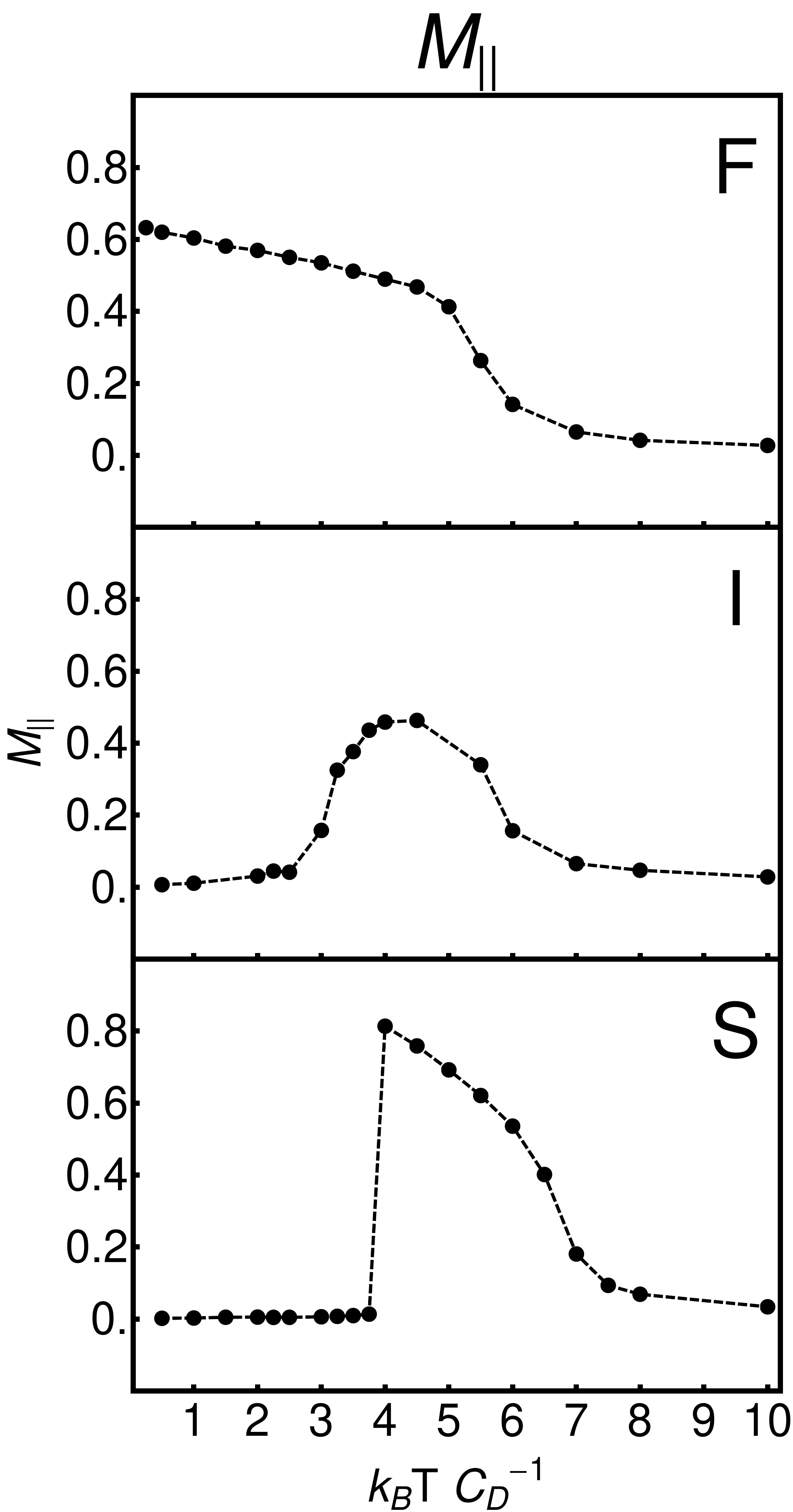}
  \caption[$M_{\parallel}$]%
  {$M_{\parallel}$ as a function of $\mathcal{T}$ for favored canting $a=-1$ (F), intermediate canting $a=0$ (I) and suppressed canting $a=2$ (S).}
  \label{MAG}
\end{figure}
\\
For the intermediate case, $a=0$, we observe results consistent with those of Whitehead et al.\cite{PhysRevB.77.174415}: a loss of orientational order over a small region, $1 < \mathcal{T} < 3$, in which stripes also become slightly canted leading to a small in-plane magnetization. Above this temperature the in-plane magnetization increases as $\eta$ decreases, resulting in a peak in-plane magnetization of $M_{\parallel} \approx .46$ at $\mathcal{T} = 4.5$. For $ \mathcal{T} > 4.5$ the $a=0$ system is identical to the $a=-1$ system, $\eta$ slowly decreasing and magnetic order gradually reduced to zero at $\mathcal{T}=8$. 
\\
For the spin suppressed state, $a=2$, we observe the same perpendicular ground state as in the $a=0$ case. Unlike the previous cases the orientational order is stabilized up to $\mathcal{T}=4$, at which point the system undergoes a sharp transition to a parallel ferromagnetic state. In Fig. \ref{CA} we see that $\eta$ simultaneously undergoes a sharp transition, representing the spin reorientation transition. Above $\mathcal{T}=4$ we observe a gradual reduction in magnetic order until the system enters the paramagnetic state above $\mathcal{T}=8$.
\section{Fluctuations}
\begin{figure}[!htb]
  \centering
  \includegraphics[width=6cm]{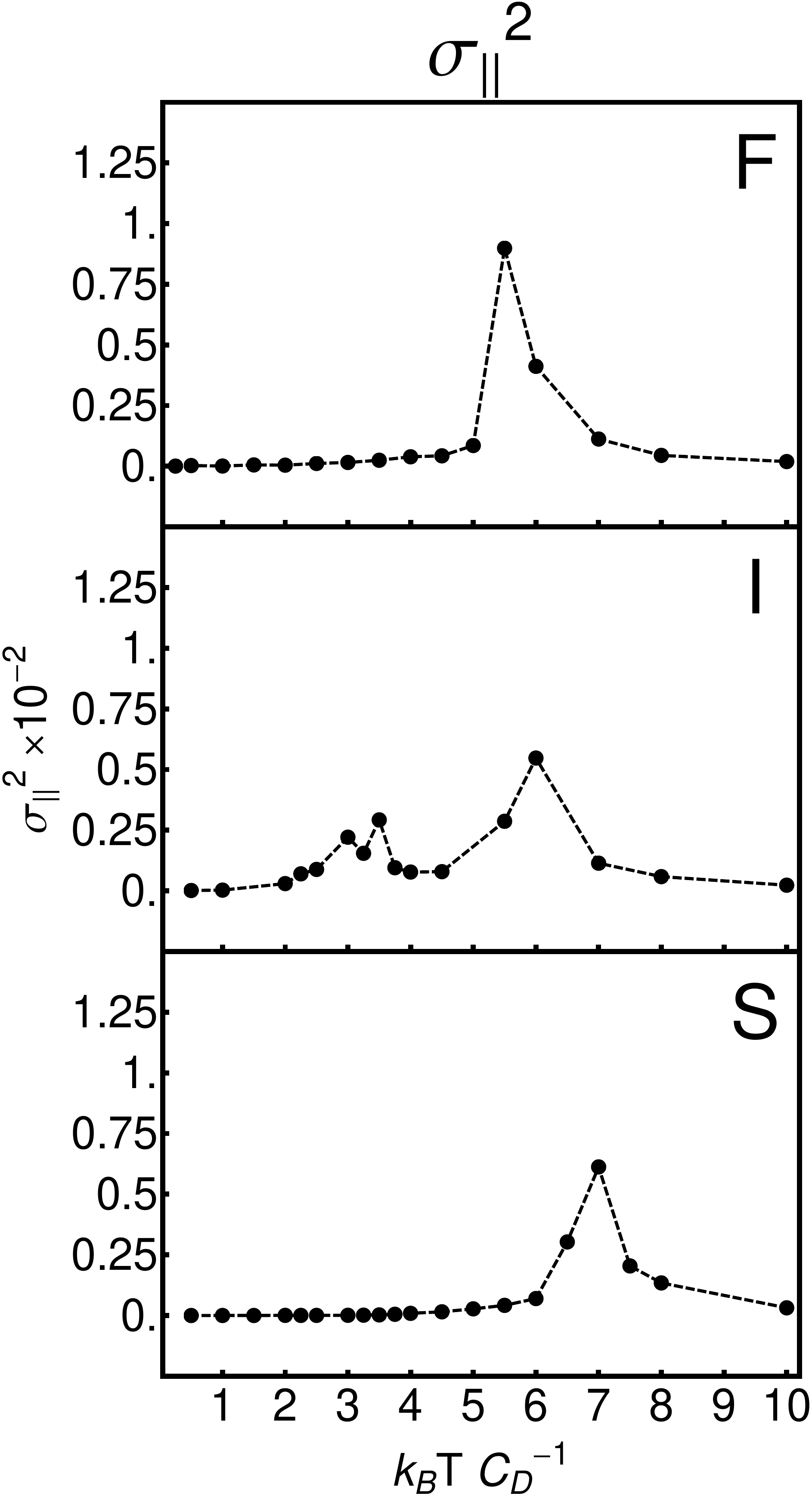}
  \caption[$\text{Var}(M_{\parallel})$]%
  {$\sigma_{\parallel}$ as a function of $\mathcal{T}$ for favored canting $a=-1$ (F), intermediate canting $a=0$ (I) and suppressed canting $a=2$ (S).}
  \label{Vmag}
\end{figure}

\begin{figure}[!htb]
  \centering
  \includegraphics[width=6cm]{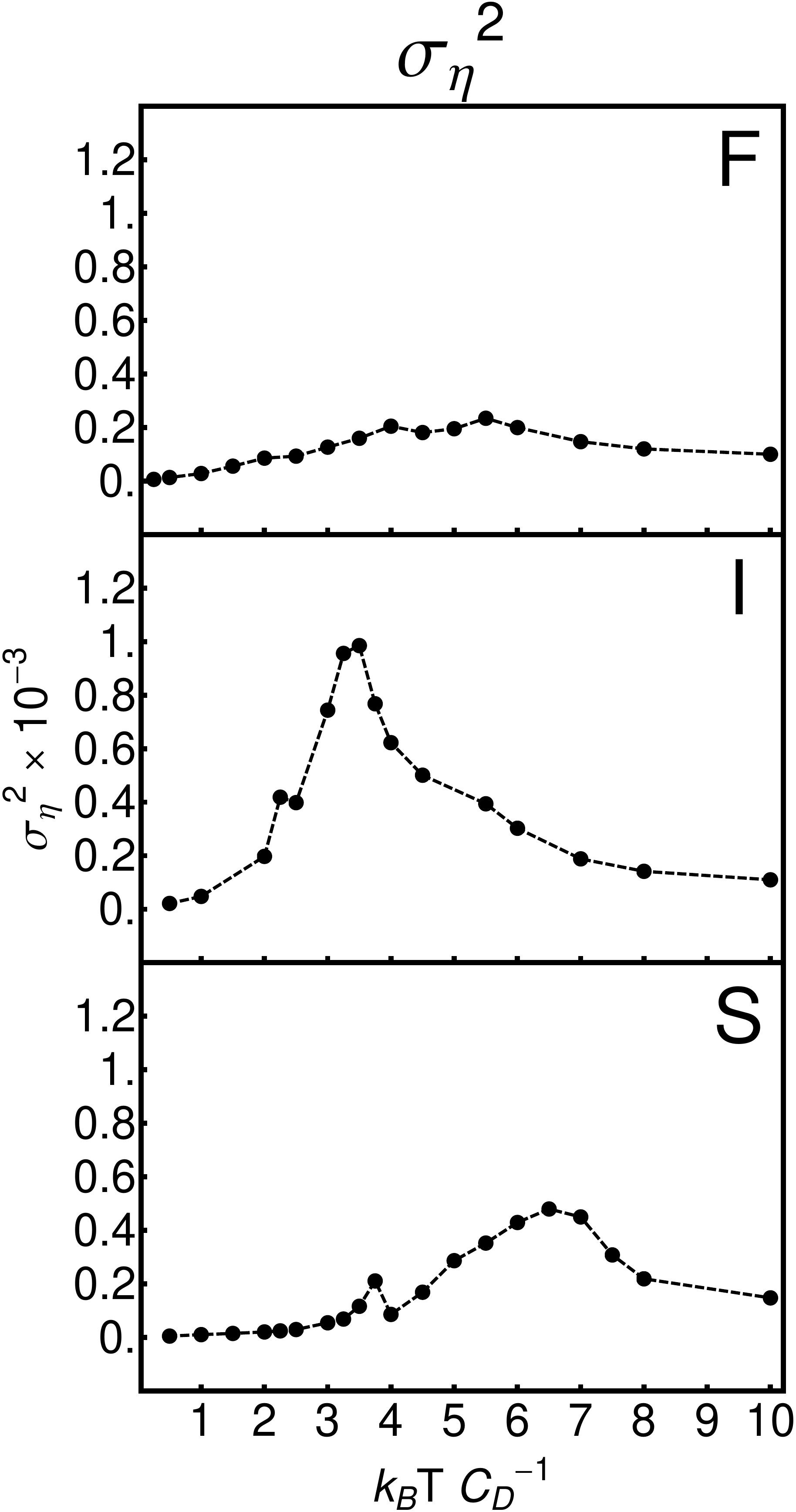}
  \caption[$\text{Var}(\eta)$]%
  {$\sigma_\eta$ as a function of $\mathcal{T}$for favored canting $a=-1$ (F), intermediate canting $a=0$ (I) and suppressed canting $a=2$ (S).}
  \label{VCA}
\end{figure}

\begin{figure}[!htb]
  \centering
  \includegraphics[width=6cm]{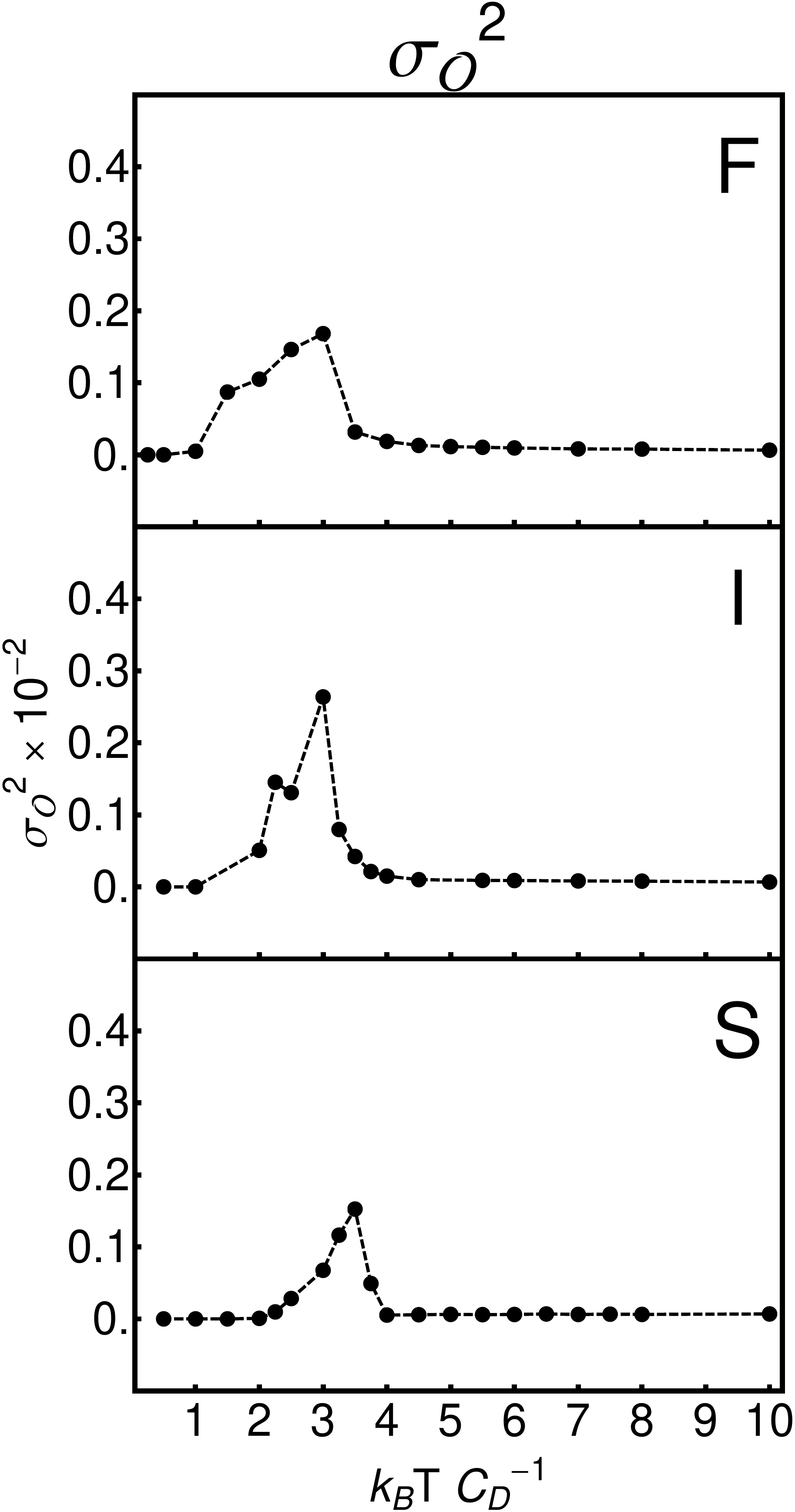}
  \caption[$\mathcal{O}^z$]%
  {$\sigma_{\mathcal{O}}$ as a function of $\mathcal{T}$ for favored canting $a=-1$ (F), intermediate canting $a=0$ (I) and suppressed canting $a=2$ (S).}
  \label{VOT}
\end{figure}

In order to examine fluctuations near critical points we calculate the auto-correlation,
\begin{equation}
\sigma^2(X) = \langle(X- \langle X \rangle)^2 \rangle,
\end{equation}
 of the three order parameters $M_{\parallel}$, $\mathcal{O}^z$ and $\eta$, which we denote $\sigma^2_{\parallel}$, $\sigma^2_{\mathcal{O}}$ and $\sigma^2_\eta$ respectively. The small numerical values of these variances means that they are affected to a greater extent by errors introduced by the simultaneous flipping. \\
In Fig. \ref{Vmag} we observe peaks in $\sigma^2_{\parallel}$ associated with the loss of in-plane magnetization for each choice of $a$. In addition we observe a smaller peak at around $\mathcal{T}=3$ for the intermediate, $a=0$, case corresponding to the formation of in-plane magnetic order. This low temperature peak is absent in the canting suppressed case due to the first order nature of the phase transition. The sharp in-plane transition does not correspond to a significant change in ferromagnetic ordering. In both the perpendicular striped phase and in-plane ferromagnetic state the exchange energy is minimized for the majority of spins. In the canting favored state the low $\mathcal{T}$ transition is absent since maximum magnetic order occurs at the lowest temperature simulated $\mathcal{T}=0.2$.
\\
In Fig. \ref{VCA} $\sigma^2_\eta$ is plotted for the three choices of $a$. For the canted favored case the fluctuations are small, culminating in shallow peak at $\mathcal{T}=5$. This is consistent with the nearly homogeneous cone angle. For the intermediate case the fluctuations displays a broad peak centered just below the minimum cone angle at $\mathcal{T}=4$. For the canting suppressed case there is a small peak corresponding to the initial reorientation of the spins followed by a broad peak as the cone angle starts to approach the high temperature average.
\\
In Fig. \ref{VOT} the fluctuations of $\mathcal{O}_z$ are plotted as a function of temperature. In each case the variance forms a peak corresponding to the loss of orienational order. Increased canting suppression corresponds to thinner peaks, as the transition occurs over a smaller temperature range. We note also that the strength of the peak is smaller for the canting suppressed case, due to the fact that only roughening occurs but not bridging. 
\section{Conclusions and Comments}
Here we have examined the nature of stripe formation and spin reorientation in the presence of strong perpendicular anisotropy. In order to do so we have  proposed an approximation that allows for a parallel algorithm for performing Monte Carlo simulations in cases where there is long range coupling. We have argued that, since the dipole coupling  contributes with a $r^{-3}$ dependence, for appropriate choice of algorithm parameters the approximation is acceptable. This algorithm reduces significantly the computation time associated with increased system size. The algorithm has been applied to the case of Ising spins in the presence of dipole coupling and shown to be consistent with results obtained by conventional methods.\\
It is possible that further increases in computational efficiency might be gained through more efficient parallelization of the dipole sum. Furthermore the concept of simultaneous flipping might be particularly useful in three dimensions where the computational issues are enhanced, the caveat of course being that care must be taken with parameter choice since the number of simultaneous spins is greatly increased. \\
This technique has then been used to examine the effects of higher order anisotropy in striped systems where we have shown that the anisotropy order can suppress or enhance the SRT in cases with strong out of plane anisotropy. \\
When comparing our simulations to analytic results we note several discrepancies. Although we were able to observe the coexistence of stripes and magnetic order, we did not observe domain formation of the cone angle as suggested by Abanov \cite{PhysRevB.51.1023}. We also did not observe any magnetic order due to correlations in the in-plane magnetization of domain walls in the wall segments dividing stripes. Whitehead et al. also noted the absence of such correlations  \cite{PhysRevB.77.174415}, however, Yafet and Gyorgy have argued that correlations between domain walls leading to ferromagnetic order are possible \cite{PhysRevB.38.9145}. The discrepancy between analytic and computational results might largely be a result of limited system size. In particular it is not possible to make a fine grained examination of wall profiles in these small systems. The parallel algorithm presented here reduces the computational cost of scaling system size. In the future it would be interesting to use the algorithm to simulate far bigger systems, in which domain wall structure could be more reasonably simulated, providing more insight into correlations that occur on finer scales.
\section*{Acknowledgements}
The authors would like to acknowledge useful discussion with Jacque Ferr\'{e} and technical assistance from Julian de Jong.This work was funded by the Australian government department of Innovation, Industry, Science and Research, the Australian Research Council, the University of Western Australia and the Scottish Universities Physics Alliance.
\appendix
\section{Pseudo Code}
\label{Code}
In order to implement the parallel algorithm described requires addressing the potential update sites and the subset of sites with which they interact as a function of the thread identifier. Here we give the pseudo code showing this addressing for the subsystems described above. The state is initialized on the host memory and consists of two one dimensional arrays of size $L^2$. These arrays store the azimuthal ($\phi$) and zenith ($\theta$) angles. In what follows we will use $\vec{s}_j$ to represent values of both $\phi$ and $\theta$ at some position $j$ in these arrays. A number of potential updates sites are selected separated horizontally and vertically by a fixed number of sites which we denote $l$, i.e $j = (al ,bl)+ j1$ . The interactions between array elements is calculated according to the algorithm below in which  blockIdx.x, blockDim.x and threadIdx.x are system integers that identify a thread address. In the following $a \backslash b$ is defined as $a\backslash b = \text{floor}( a/b )$. For example $7\backslash 3 = 2$ and $2 \times (7\backslash 3) + 7 \text{mod} 3 = 7$.
\begin{algorithmic}
 \STATE  \bf{ON HOST}
		\STATE	$S_1$ = Random Integer $\in [1,l]$ 
		\STATE	$S_2$ = Random Integer $\in [1,l]$ 
		\STATE        $N = L\backslash l$
		\FOR {$j=1$ ; $j < N$ }
		\STATE Select a series of new states:
		\STATE 	$\theta[j] $ = Random Real $\in [0,2 \pi]$
		\STATE  	$\phi[j]   $ = Arccos(Random Real $\in [-1,1]$)
		\STATE Copy all variables arrays to GPU
		\ENDFOR
\end{algorithmic}
\begin{algorithmic}
 \STATE  \bf{ON GPU: Exchange Interaction}		
	\STATE $T_\text{id}$ = blockIdx.x*blockDim.x + threadIdx.x;
	\IF	{$T_\text{id}\leq L^2 \backslash l^2 $}
	
	\STATE $S_x=S_1 + l(T_\text{id} \text{mod} N)$
	\STATE $S_y=(S_2+l(T_\text{id} \backslash N))$
	\STATE $j = S_yL+S_x$
	\STATE calculate the exchange coupling between $\vec{s}_j$ and its immediate neighbors
	\STATE calculate the exchange coupling between $(\theta[j],\phi[j])$ and $\vec{s}_j$'s immediate neighbors	
	\ENDIF	
\end{algorithmic}
\begin{algorithmic}
 \STATE  \bf{ON GPU: Dipole Interaction}

	\STATE $T_\text{id}$ = blockIdx.x*blockDim.x + threadIdx.x
	\STATE $S_\text{id}$ = $T_\text{id} \backslash P$
	
	\IF	{$S_\text{id}\leq L^2 \backslash l^2 $}
	
	\STATE $S_x=S_1 + l(S_\text{id}\text{mod} N)$
	\STATE $S_y=(S_2+l(S_\text{id} \backslash N))$
	\STATE $j = S_yL+S_x$

	\STATE $\sigma$ = $(L+1) \backslash P$
	\STATE $\rho = ((L+1)\text{mod}P)-1$
	\FOR {$k=-L\backslash2$ ; $k\leq L \backslash 2$ }
	\FOR {$i=-L \backslash2 + \sigma T_\text{id} \text{mod}P)$ ;	$i < -L  \backslash 2 + \sigma(T_\text{id}\text{mod} P+1) $  }
	
	\STATE $j' = (((S_y+k)\text{mod}L)L)+(S_x+i)\text{mod}L$ 
	\STATE Calculate the dipole coupling between $\vec{s}_j$ and $\vec{s}_{j'}$
	\STATE Calculate the dipole coupling between $(\theta[j],\phi[j])$ and $\vec{s}_{j'}$
	\ENDFOR
	\ENDFOR
	\ENDIF
\end{algorithmic}
\begin{algorithmic}
 \STATE  \bf{ON GPU: Single Site Energies and Spin Flips}		
	\STATE $T_\text{id}$ = blockIdx.x*blockDim.x + threadIdx.x;
	\IF	{$T_\text{id}\leq L^2 \backslash l^2 $}
	
	\STATE $S_x=S_1 + l(T_\text{id} \text{mod} N)$
	\STATE $S_y=(S_2+l(T_\text{id}  \backslash N))$
	\STATE $j = S_yL+S_x$
	\STATE calculate anisotropy and Zeeman energies for $\vec{s}_j$
	\STATE calculate anisotropy and Zeeman energies $(\theta[j],\phi[j])$	
	\STATE replace $\vec{s}_j$ with $(\theta[j],\phi[j])$ or $\vec{s}_j$ according to the Boltzmann probability	
	\ENDIF	
\end{algorithmic}
\section{Dipole Ising Model}
\label{DIM}
In section \ref{SectA} it was argued that the error introduced by a single pass of the proposed GPU algorithm is bounded. However this does not ensure that the error is not compounded over a large number of passes leading to large systematic errors. In order to investigate this possibility we consider the result of applying the algorithm when every accepted spin update corresponds to a complete reversal: $\vec{s}_i = (0,0,s_i^z) = (0,0,\pm 1)$. We select the same algorithm parameters that are used in section \ref{Res}: $L=64$ and $l=32$. This corresponds to four simultaneous attempted spin flips for each cycle of the algorithm. Since each site is restricted to only two states the energy can be written
\begin{equation}
 H= \frac{J}{2}\sum_{\langle i,j \rangle} s_i  s_j +\frac{C_D}{2} \sum_{i,j} \frac{s_i s_j}{r_{ij}^3} 
\end{equation}
 In the presence of a sufficiently strong dipole interaction the ground state of the system will form a striped pattern of alternating spins. In order to ensure that dipole coupling contributes significantly to the total energy (increasing the error)  parameters are selected such that $J/C_D = 1.7$, which corresponds to the thinnest possible stable stripe width $h=1$. The order parameter for such a system is formed by considering a series of sub-lattices in the manner described by Binder and Landau \cite{PhysRevB.21.1941}. For $h=1$ the system is broken into four sub-lattices $m_\lambda$,horizontal stripes are described by $m_h = m_1+m_2-m_3-m_4$ and vertical stripes by $m_v = m_1+m_4-m_3-m_2$. The order parameter is then the staggered magnetization $m_{\text{st}} = \langle (m_h^2 + m_v^2 )^{1/2} \rangle$. In Fig. 17 
this staggered magnetization is shown as a function of a normalized temperature $\mathcal{T}= k_B T C_D^{-1}$.
\begin{figure}[!htb]
  \centering
  \includegraphics[width=6cm]{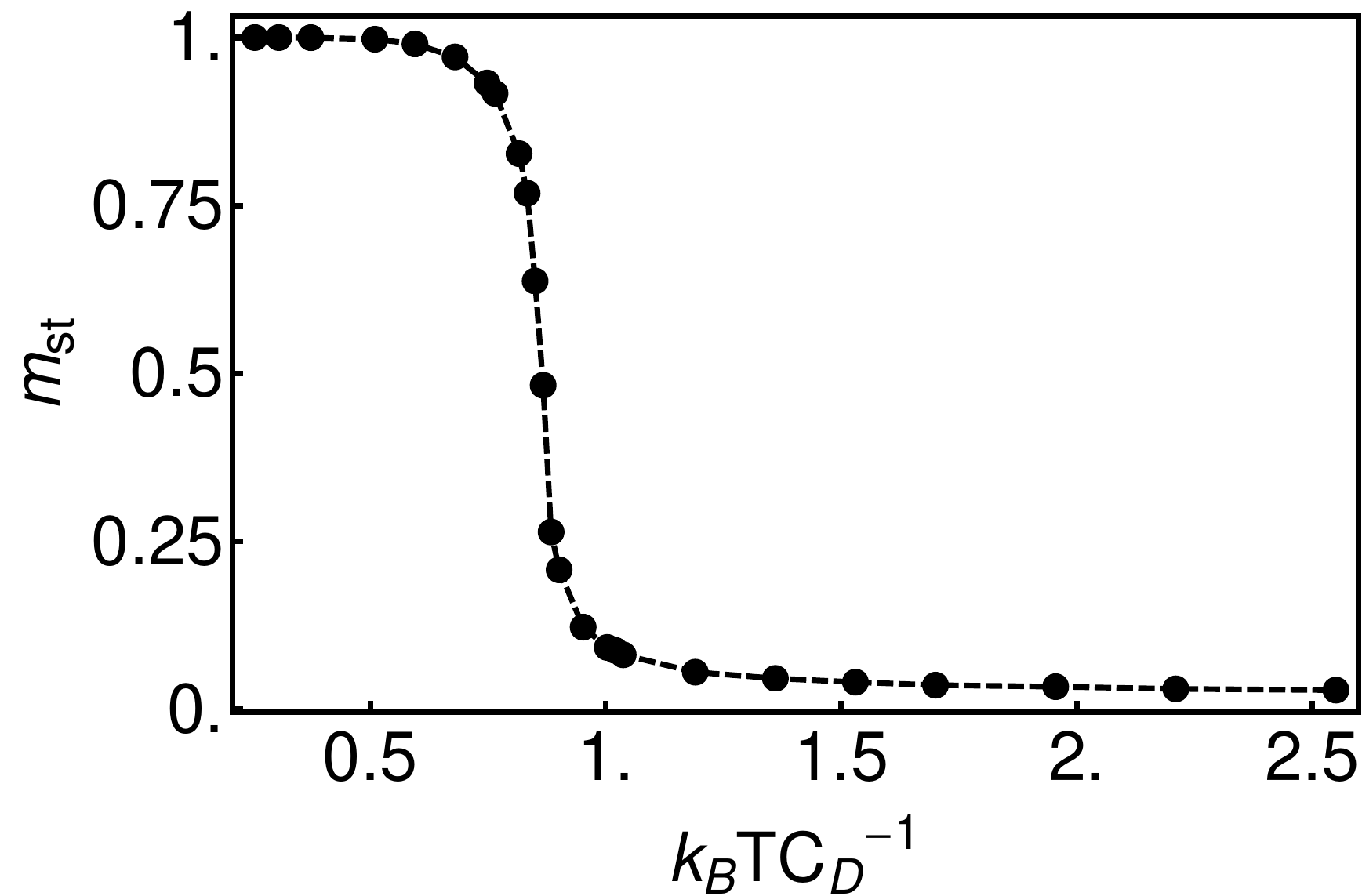}
  \caption[Staggered Magnetization]%
  {The staggered magnetization as a function of $\mathcal{T}$ showing the order disorder transition.}
  \label{mst}
\end{figure}
\begin{figure}[!hb]
  \centering
  \includegraphics[width=6cm]{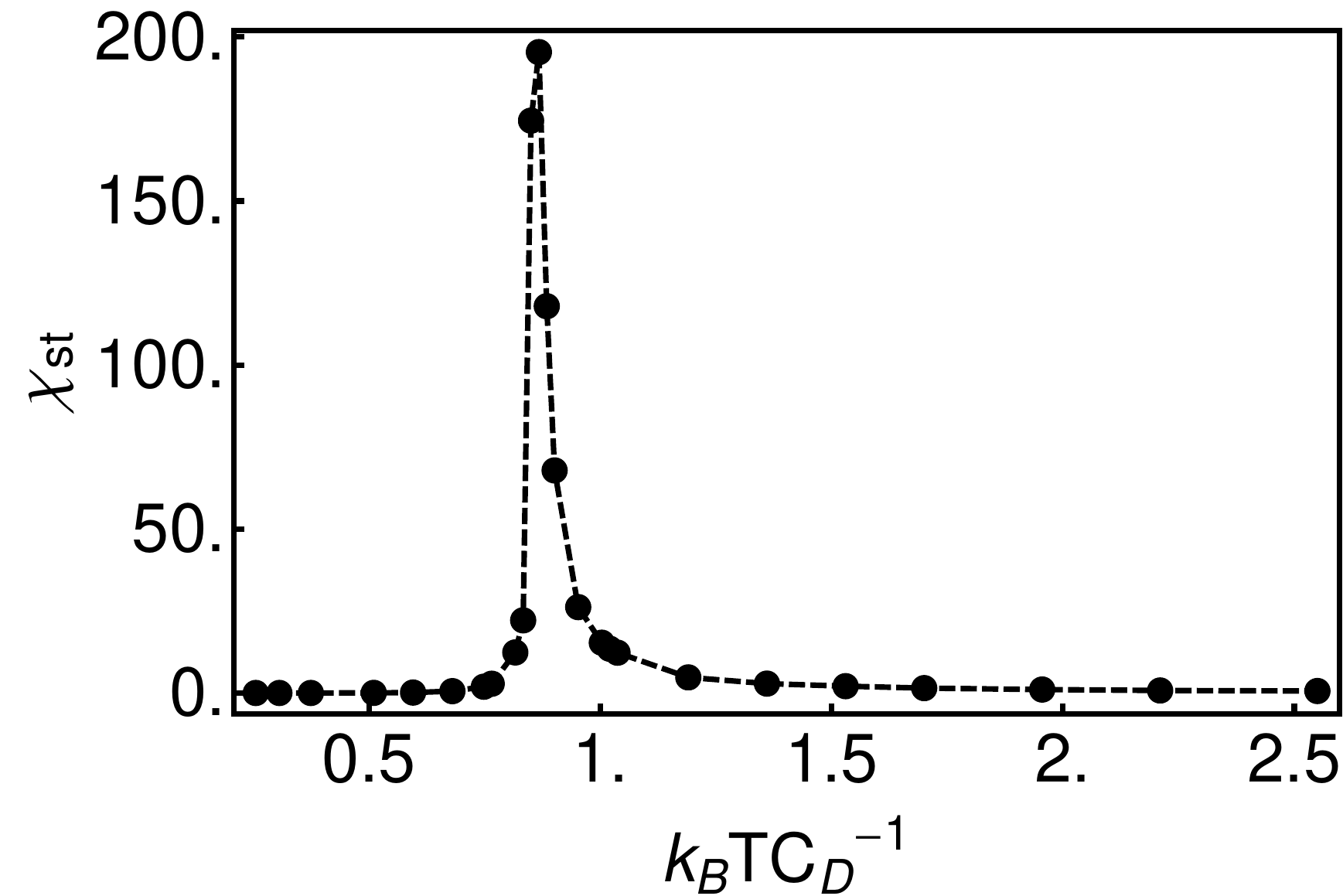}
  \caption[Staggered Susceptibility]%
  {The staggered Susceptibility as a function of $\mathcal{T}$.}
  \label{msus}
\end{figure}
The staggered susceptibility $\chi_{\text{st}} = \frac{L^2}{k_B T} (\langle m_{\text{st}}^2 \rangle -\langle m_{\text{st}} \rangle ^2)$ is also calculated. By determining the location of the peak in Fig. 
18 the critical temperature of the phase transition $\mathcal{T}_c$ can be determined, the result is shown in Table I.
 The peak lacks the $\delta$-like structure associated with a first order transition, instead displaying an exponential decay consistent with a continuous phase transition \cite{PhysRevB.34.1841,PhysRevB.69.092409}, this is consistent with previous simulations with strong dipole coupling \cite{engdist}. In Fig. 
19 we show the probability distribution of the average energy per spin for various temperatures near the transition. At each temperature the distribution displays a single turning point, this is also  consistent with the expected continuous transition \cite{engdist}.\\
\begin{figure}[!hb]
  \centering
  \includegraphics[width=7cm]{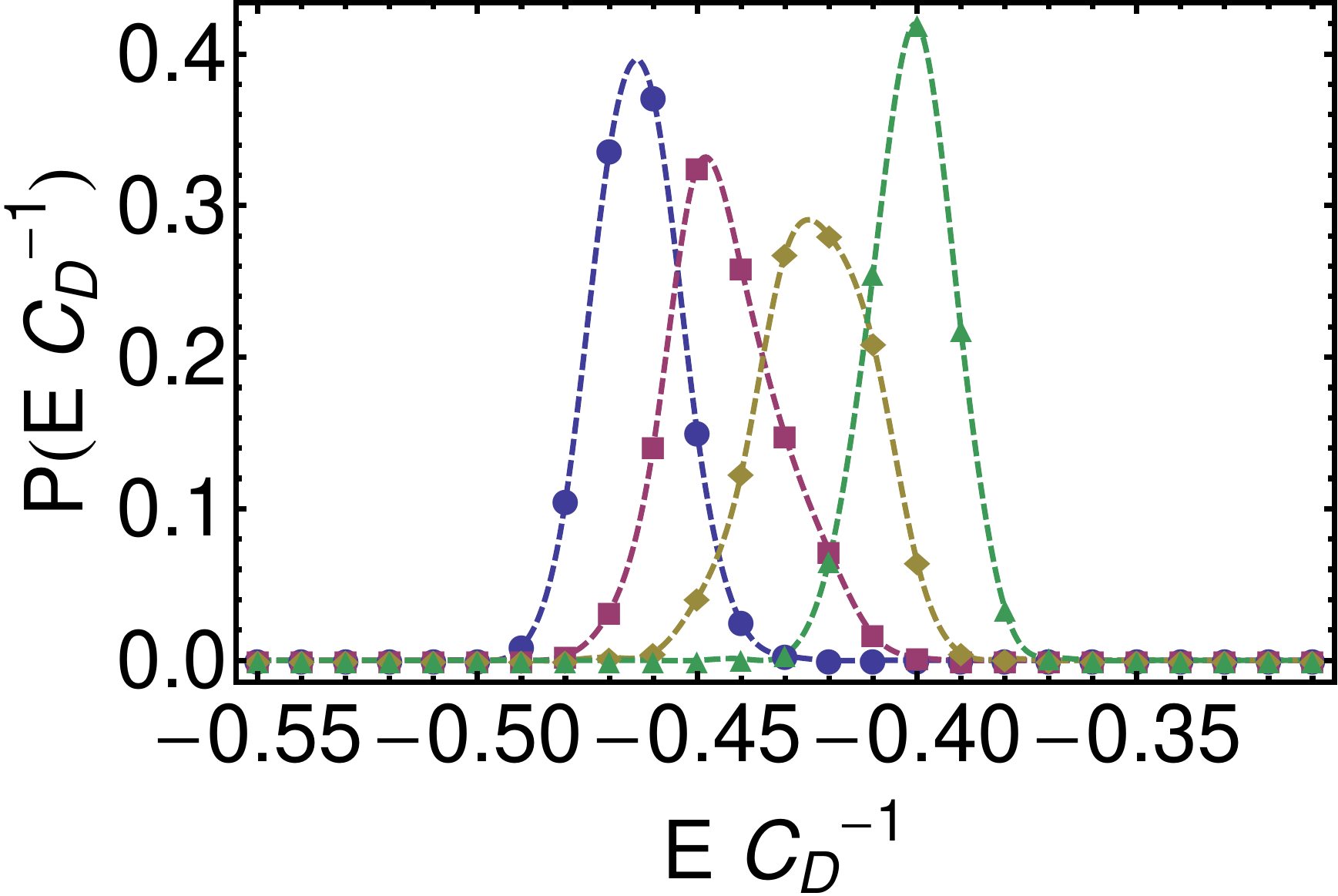}
  \caption[Probability distributions of energy near the critical point]%
  {Probability distributions of energy near the critical point:
  $\mathcal{T}= .833$ Blue Circles;  $\mathcal{T}= .85$ Red Squares; $\mathcal{T}= .867$ Yellow Diamonds; $\mathcal{T}= .884$ Green Triangles.}
  \label{Short}
\end{figure}
Near to the critical temperature the system displays critical behavior. After rescaling temperature as $t= L^{\frac{1}{\nu}} \lvert 1- T/T_c \rvert  $ one expects the following scaling relationships: $m_{\text{st}} = L^{-\frac{\beta}{\nu}} m_0 (t) $ and $\chi_{\text{st}}= \frac{L^{\frac{\gamma}{\nu}}}{t} \chi_0 (t) $, where $m_0 (t)$ and $\chi_0 (t)$ are universal scaling functions \cite{Stats, PhysRevB.73.144418}. Close to the critical point, when $\lvert 1- T/T_c \rvert$ is small and $L$ large, these universal scaling functions are expected to show a power law behavior $m_0 (t) \rightarrow B t^{ \beta} $ and $\chi_0 (t) \rightarrow A t^{-\gamma}$. In order to extract parameters from simulation one usually makes use of the universality of  $m_0 (t)$ and $\chi_0 (t)$ and simulates the system for a number of different sizes before varying the parameters until the results of all simulations lie on a single curve \cite{crit_234dising_zphysb1981,PhysRevB.73.144418}. However, performing such analysis would not be a suitable test of the algorithm, for any choice $L<l$ there is no simultaneous flipping and hence no approximation is being made. If one were to change the size of $l$  to simulate smaller systems the degree of approximation would be changed (the error will be increased for decreasing $l$). Instead we use the value of the scaling constant $\nu$ calculated by Rastelli et al. to rescale temperature and then use the power law dependence of the system near $T_c$ to extract $\beta$ and $\gamma$. A comparison of the critical properties is given in Table I. 
\begin{table}[htb]
\caption{Various parameters calculated using the GPU algorithm and conventional techniques \cite{PhysRevB.73.144418}}
\begin{tabular}{ccc}
\hline\hline
Parameter & GPU algorithm & Conventional MC \\
\hline
$\mathcal{T}_c$ & .85 &.82\\
$\gamma$ & 1.62 &1.75 \\
$\beta$ & 0.08 &0.08 \\
\hline
\end{tabular}
\label{res}
\end{table}
The use of the GPU has introduced some error in the critical parameters of the system, the largest error being slightly less than $8\%$. Previous simulations performed on GPUs have noted error averaging $5\%$ in the calculation of dipole energy due to the low level of numerical precision available on GPUs \cite{campos}. In light of this reduced precision the additional error introduced by multiple spin flips is acceptable.

\bibliographystyle{unsrt}	
\bibliography{PHM}
\end{document}